\documentclass[10pt, conference]{IEEEtran}
\makeatletter  
\newif\if@restonecol  
\makeatother

\usepackage[linesnumbered,ruled,vlined]{algorithm2e}%[ruled,vlined]{  
\usepackage{algpseudocode}  
\usepackage{amsmath}  
\usepackage{amssymb}
\usepackage{cite}
\usepackage{multirow} 
\usepackage{graphicx}
\usepackage{subfigure}
\usepackage{subfig}
\usepackage{booktabs}
\DeclareMathOperator*{\argmin}{arg\,min}

\renewcommand{\baselinestretch}{0.985}
\begin{document}
%
% paper title
% Titles are generally capitalized except for words such as a, an, and, as,
% at, but, by, for, in, nor, of, on, or, the, to and up, which are usually
% not capitalized unless they are the first or last word of the title.
% Linebreaks \\ can be used within to get better formatting as desired.
% Do not put math or special symbols in the title.

\title{\LARGE \bf FedRAV:  Hierarchically Federated Region-Learning for Traffic Object Classification of Autonomous Vehicles}

\author{\IEEEauthorblockN{
Yijun Zhai\IEEEauthorrefmark{2}, Pengzhan Zhou\IEEEauthorrefmark{2}, Yuepeng He\IEEEauthorrefmark{2}, 
Fang Qu\IEEEauthorrefmark{2}, 
Zhida Qin\IEEEauthorrefmark{4},
Xianlong Jiao\IEEEauthorrefmark{2}, 
Guiyan Liu\IEEEauthorrefmark{2}, and 
Songtao Guo\IEEEauthorrefmark{2}
}
\IEEEauthorblockA{\IEEEauthorrefmark{2}College of Computer Science, Chongqing University, Chongqing, China,\\ \IEEEauthorrefmark{4}School of Computer Science and Technology, Beijing Institute of Technology, Beijing, China}
Email: yjzhai@stu.cqu.edu.cn and pzzhou@cqu.edu.cn
}

% \lipsum[10] \cite{math_65_source_code}

% \lipsum[10]

% \IEEEauthorblockA{\IEEEauthorrefmark{2}Twentieth Century Fox, Springfield, USA\\
% Email: homer@thesimpsons.com}

% \IEEEauthorblockA{\IEEEauthorrefmark{3}Starfleet Academy, San Francisco, California 96678-2391\\
% Telephone: (800) 555--1212, Fax: (888) 555--1212}

% \IEEEauthorblockA{\IEEEauthorrefmark{4}Tyrell Inc., 123 Replicant Street, Los Angeles, California 90210--4321}}

% use for special paper notices
%\IEEEspecialpapernotice{(Invited Paper)}

% make the title area
\maketitle

% As a general rule, do not put math, special symbols or citations
% in the abstract

\begin{abstract}
The emerging federated learning enables distributed autonomous vehicles to train equipped deep learning models collaboratively without exposing their raw data, providing great potential for utilizing explosively growing autonomous driving data. However, considering the complicated traffic environments and driving scenarios, deploying federated learning for autonomous vehicles is inevitably challenged by non-independent and identically distributed (Non-IID) data of vehicles, which may lead to failed convergence and low training accuracy. In this paper, we propose a novel hierarchically Federated Region-learning framework of Autonomous Vehicles (FedRAV), a two-stage framework,which adaptively divides a large area containing vehicles into sub-regions based on the defined region-wise distance, and achieves personalized vehicular models and regional models. This approach ensures that the personalized vehicular model adopts the beneficial models while discarding the unprofitable ones. We validate our FedRAV framework against existing federated learning algorithms on three real-world autonomous driving datasets in various heterogeneous settings. The experiment results demonstrate that our framework outperforms those known algorithms, and improves the accuracy by at least 3.69\%. The source code of FedRAV is available at: https://github.com/yjzhai-cs/FedRAV.

\end{abstract}


% \begin{IEEEkeywords}
% Federated Learning, Hypernetwork, Partitioning Method.
% \end{IEEEkeywords}

% no keywords

% For peer review papers, you can put extra information on the cover
% page as needed:
% \ifCLASSOPTIONpeerreview
% \begin{center} \bfseries EDICS Category: 3-BBND \end{center}
% \fi
%
% For peerreview papers, this IEEEtran command inserts a page break and
% creates the second title. It will be ignored for other modes.
\IEEEpeerreviewmaketitle

\section{Introduction} \label{intro}
% Autonomous Electric Vehicles (AEVs)
Autonomous Vehicles (AVs)
have made significant progress in industry and academia in recent years. Advanced autonomous driving technology liberates humans from laboring driving experiences while also raising numerous research challenges and open questions. AVs are required to address high-dimensional scenarios, the combinations of vehicles, traffic lights, pedestrians, objects, and weather conditions, which seem rarely satisfied with the amount of fed training data. As a prospective resolution, federated learning (FL) \cite{mcmahan2017communication,fallah2020personalized,sattler2020clustered,vahidian2023efficient} incorporates clients to train a global model using the data collected separately and preserves clients' privacy.

% A standard FL system first selects some available clients and sends them a global model. Then, each selected client performs stochastic gradient descent (SGD) on the local dataset and uploads the model parameter or gradient to the parameter server. Finally, the parameter server can aggregate these models into a holistic model and treat it as the
% % [zz]
% initial model for the next communication round.

However, deploying FL for AVs is in the face of vehicles' non-independent and identically distributed data (i.e. Non-IID, also known as heterogeneous data) \cite{zhao2018federated,li2020federated,wang2020tackling}, especially considering the complicated traffic environments and driving scenarios. For example, the data collected from highways, urban areas, rural areas, schools, coasts, and mountains vary significantly in their distributions.
Intuitively, data acquired from a smaller region tend to have more similar distributions, such as the data collected from the vehicles passing the same traffic light tend to be similar. Based on this observation, a 
% \textbf{F}ederated \textbf{R}egion-\textbf{L}earning framework (FRL) 
hierarchically \underline{Fed}erated \underline{R}egion-learning framework of \underline{A}utonomous \underline{V}ehicles (FedRAV)
is proposed in this paper to establish an FL framework to learn from vehicles in sub-regions of a much larger area.
Realizing FedRAV can help AVs learn autonomous driving models based on data of a more IID distribution, where many challenges need to be addressed.
%which inevitably faces many challenges. 
For example, \textit{how should a large area be divided into sub-regions adaptively considering the data distribution? } To address the challenge, a partitioning mechanism with one-shot communication is proposed in this paper, which divides the area of interest into sub-regions of different sizes based on the region-wise distance, achieving better IID data in each region.
% [zz]%把全文的sub xx换成subxx，比如sub region写成subregion，这是标准写法。

Furthermore, \textit{how should the system learn a specialized model for each AV to fit its local driving data?}
%local model parameter space on highly Non-IID driving data? }
In order to specialize the models learned in each AV, a personalization strategy via hypernetworks is proposed. FedRAV employs a designated hypernetwork to learn specialized \textit{mask vectors} per vehicle and personalizes the vehicular model by using the mask vector to weight the models shared by vehicles within the same region. 
% used in the linear combination of models shared by vehicles within the same region. 
These mask vectors ensure that the specialized vehicular model adopts the beneficial models while discarding the unprofitable ones.
% These mask vectors strike a balance between facilitating knowledge transfer and mitigating the influence of heterogeneous model space. 
% [zz]%没看懂
Furthermore, the same hypernetwork design is also applied to different regions, which allows vehicles to learn a personalized vehicular model while simultaneously benefiting from the specialized regional model based on regional data. Therefore, the proposed framework can address the Non-IID problem of AVs' data and train the learning model based on numerous vehicles navigating in different regions, which promotes better autonomous driving adapted for driving scenarios.
%elevates better user experience.

The contributions of the paper are mainly three-fold:
\begin{itemize}
\item We empirically demonstrate that the spatial distribution of vehicles' collected data exhibits regional similarities and design a partitioning mechanism 
% with guaranteed approximation bound 
to divide large driving areas into sub-regions effectively.

\item We propose a novel hierarchically Federated Region-learning framework of Autonomous Vehicles suitable for achieving federated learning of vehicles considering the Non-IID driving data. Our framework presents an effective personalization strategy, exploiting richer data of plentiful vehicles to train better personalized vehicular models while maintaining regional models.
% [zz]%exploiting the broad scope of data from numerous vehicles 
% while maintaining the personalized vehicular models and regional models. 

\item We evaluate the performance of FedRAV on three real-world autonomous driving datasets and open-access the source code of FedRAV to contribute to the community.
% The simulation results exhibit that FedRAV improves final accuracy by $+8.7\%$ on GTSRB \cite{Houben-IJCNN-2013}, $+3.69\%$ on MIO-TCD \cite{luo2018mio} and $+9.36\%$ on Vehicle-10 for LG-FedAvg \cite{liang2020think}.

% final accuracy by $+11.93\%$ on GTSRB \cite{Houben-IJCNN-2013}, $+0.57\%$ on MIO-TCD \cite{luo2018mio} and $+5.91\%$ on Vehicle-10 for PACFL \cite{vahidian2023efficient}.

% \item In addition, we open the source code of FedRAV and vehicle classification dataset Vehicle-10\footnote{Vehicle-10 dataset is open sourced at: anonymous link. } to contribute to the community.

% \item In additation, we open source code of FRL and vehicle classification dataset Vehicle-10 to contribute to the community.

\end{itemize}

% [zz]
%加一段这个，The remained paper are organized as following. Section \ref{sec:prelim} XX.

\begin{figure*}[!ht]
\subfigure{
    \begin{minipage}[h]{0.32\linewidth}
        \subfigure{
            \includegraphics[width=1.1in]{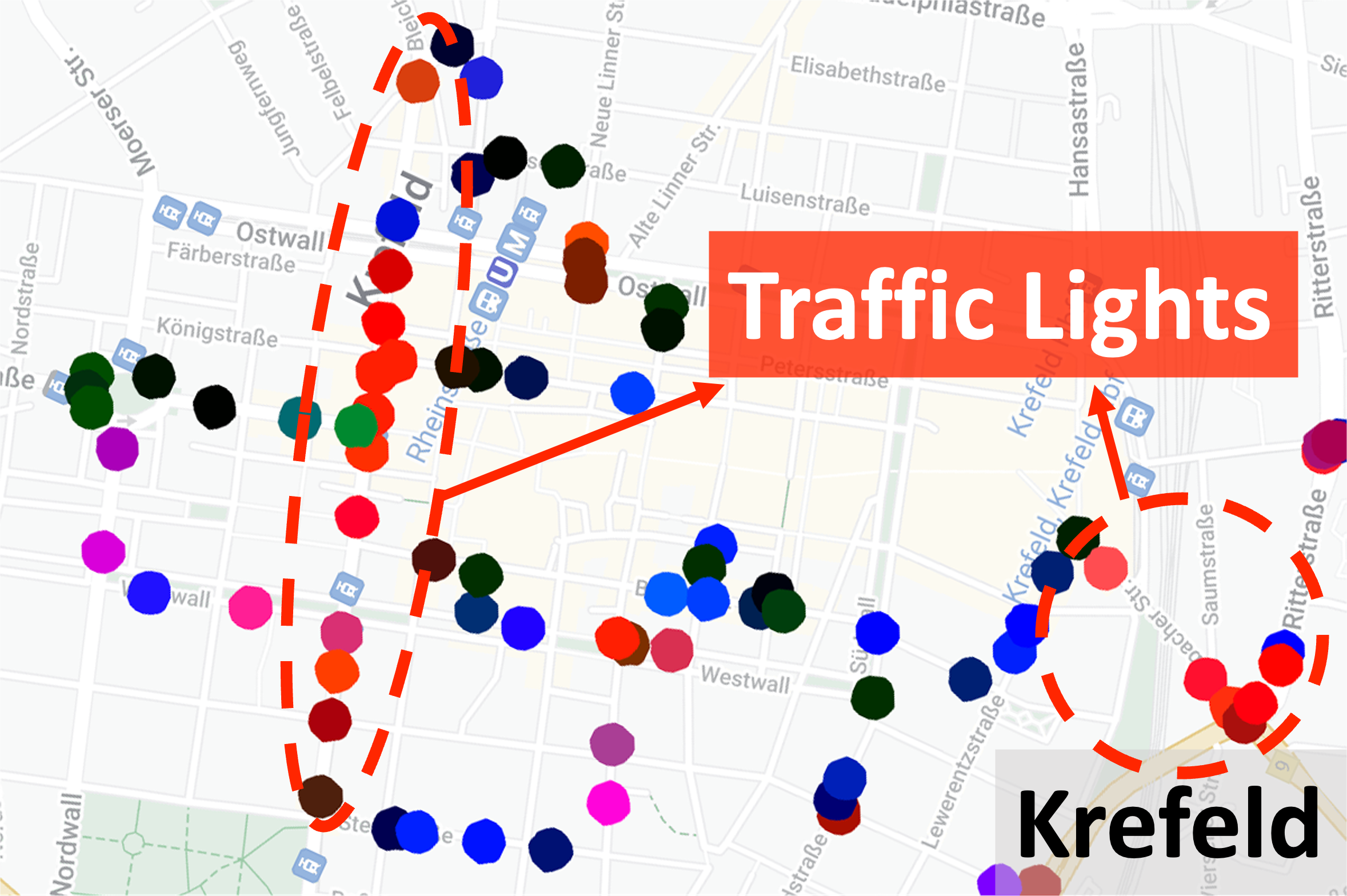} \\ \hspace{0.025in}
            \includegraphics[width=1.1in]{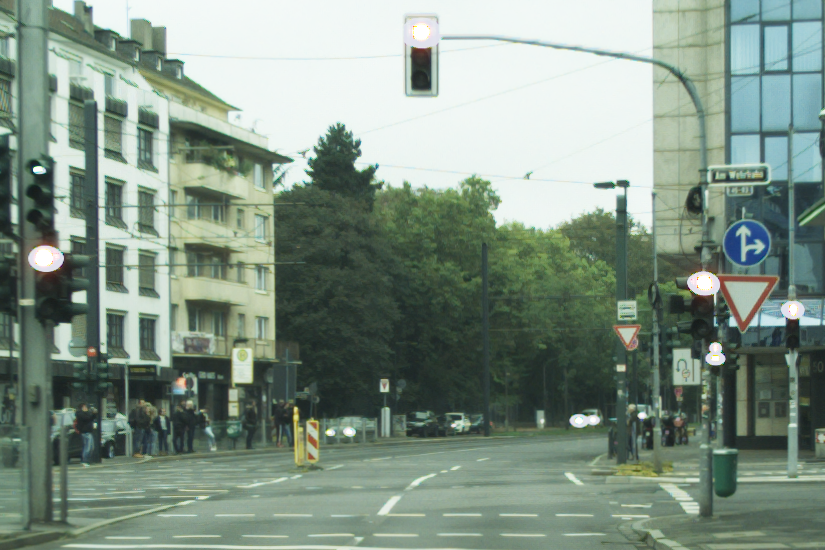} \\ \hspace{0.025in}
        }
        \subfigure{
            \includegraphics[width=1.1in]{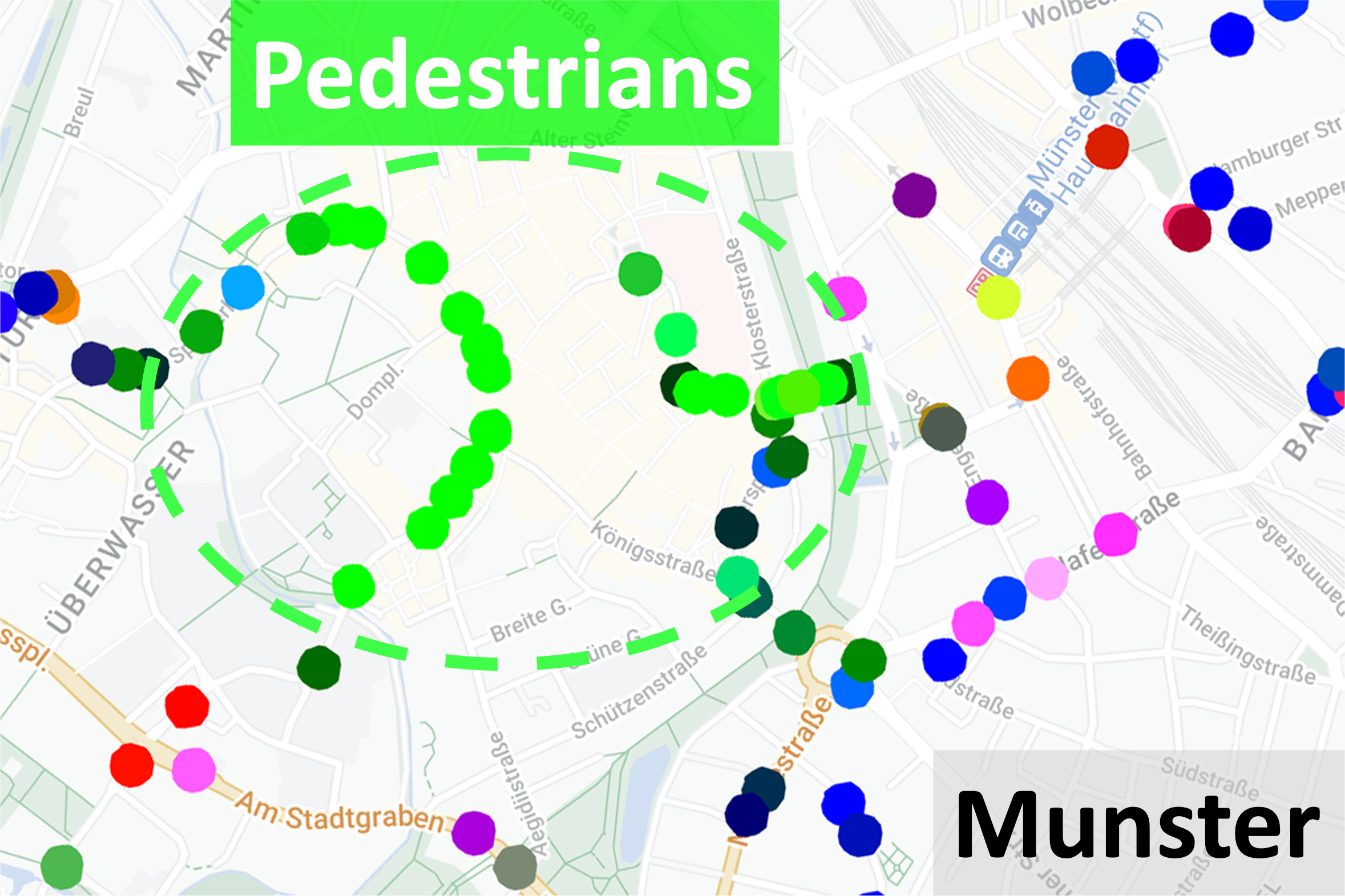} \\ \hspace{0.025in}
            \includegraphics[width=1.1in]{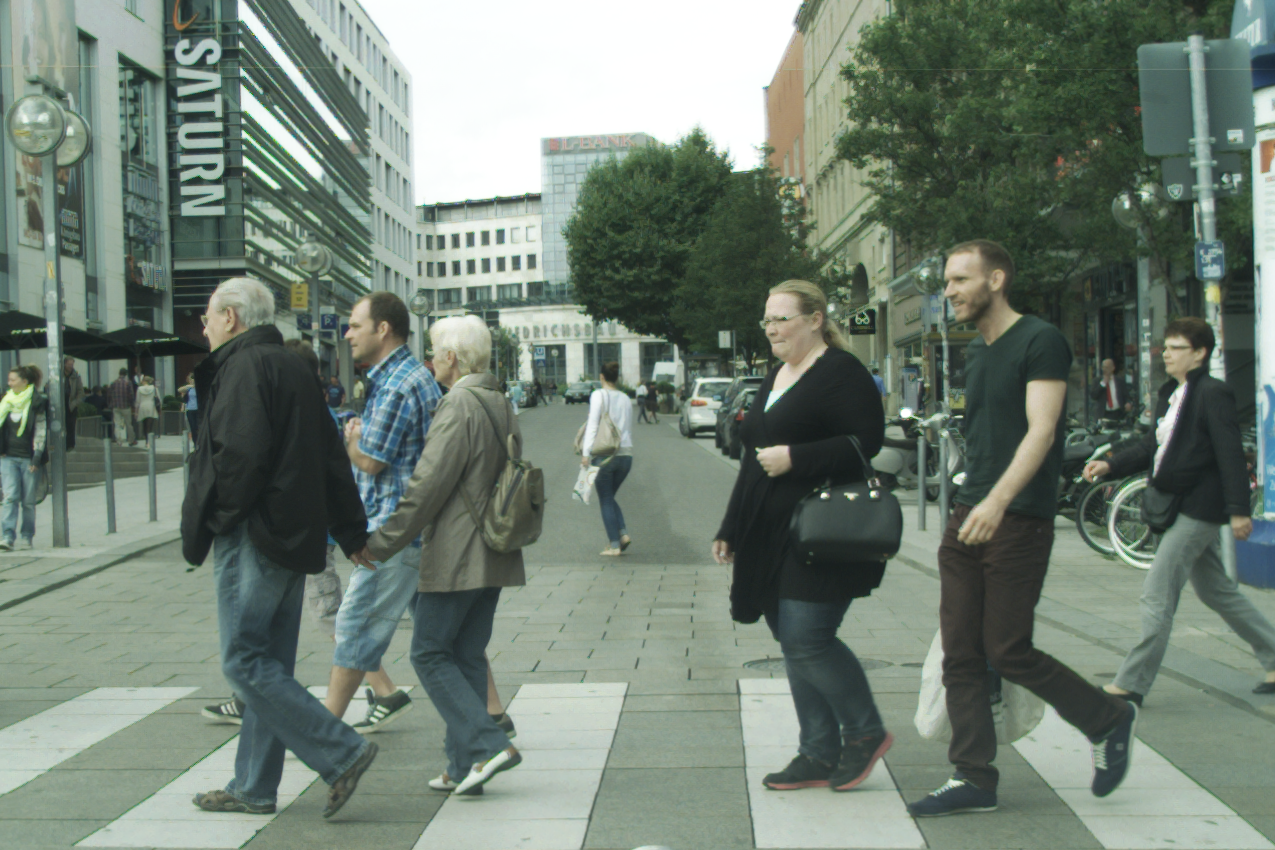} \\ \hspace{0.025in}
        }
        \subfigure{
            \includegraphics[width=1.1in]{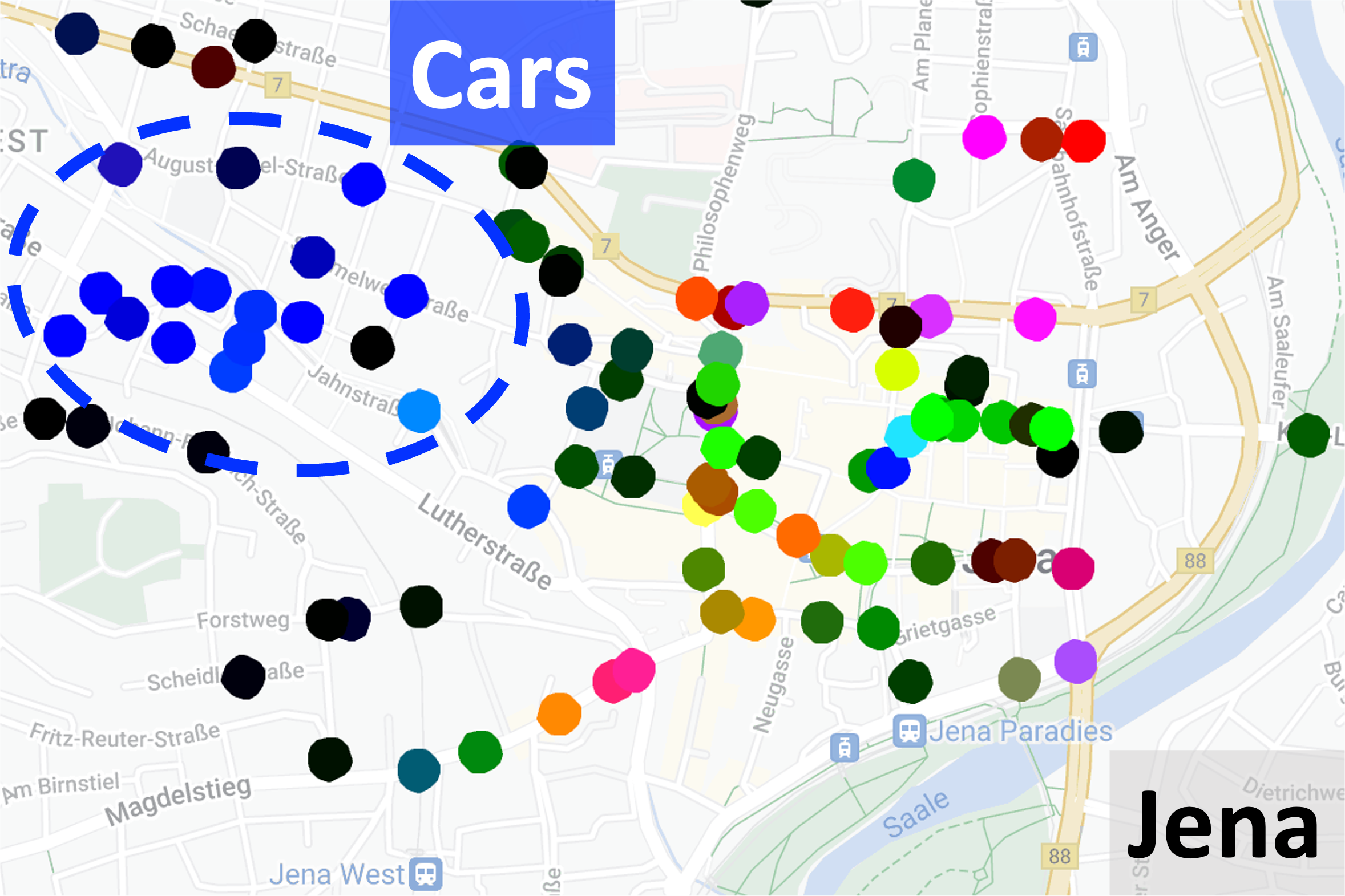} \\ \hspace{0.025in}
            \includegraphics[width=1.1in]{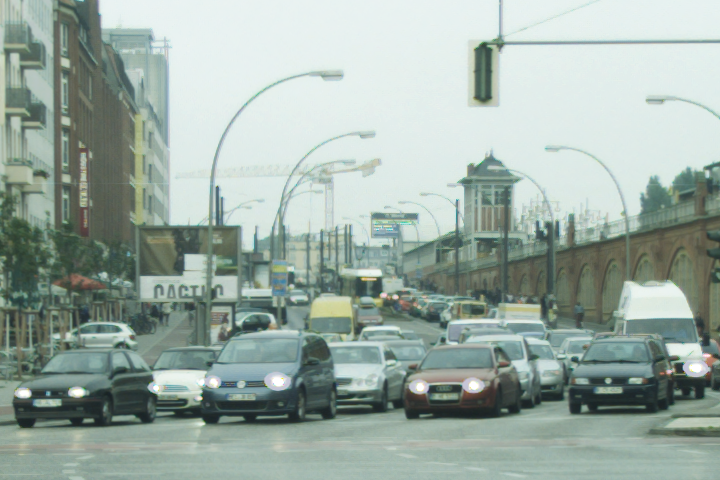} \\ \hspace{0.025in}
        }
    \end{minipage}
    \begin{minipage}[h]{0.32\linewidth}
        % \centering
        \subfigure{
            \includegraphics[width=1.1in]{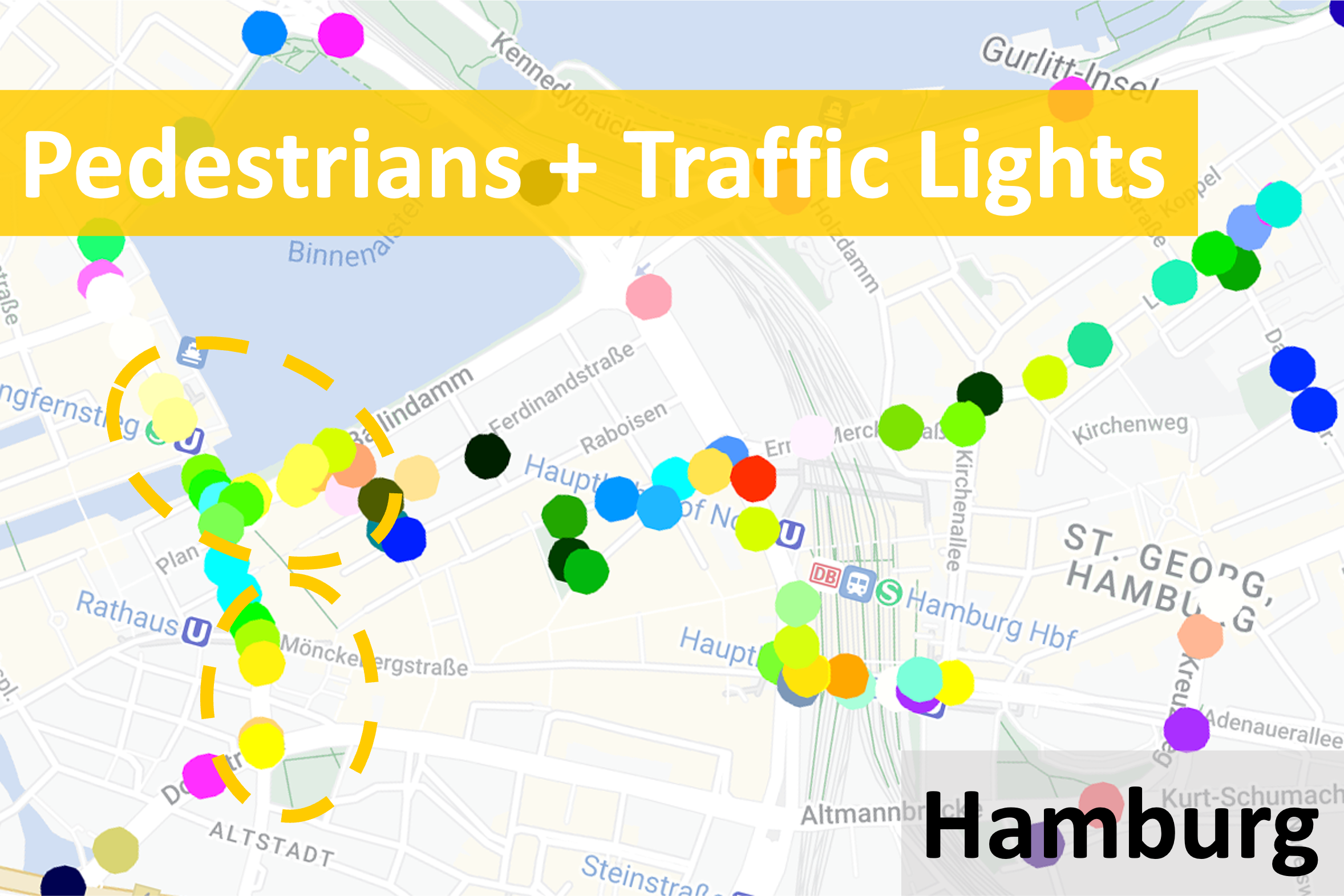} \\ \hspace{0.025in}
            \includegraphics[width=1.1in]{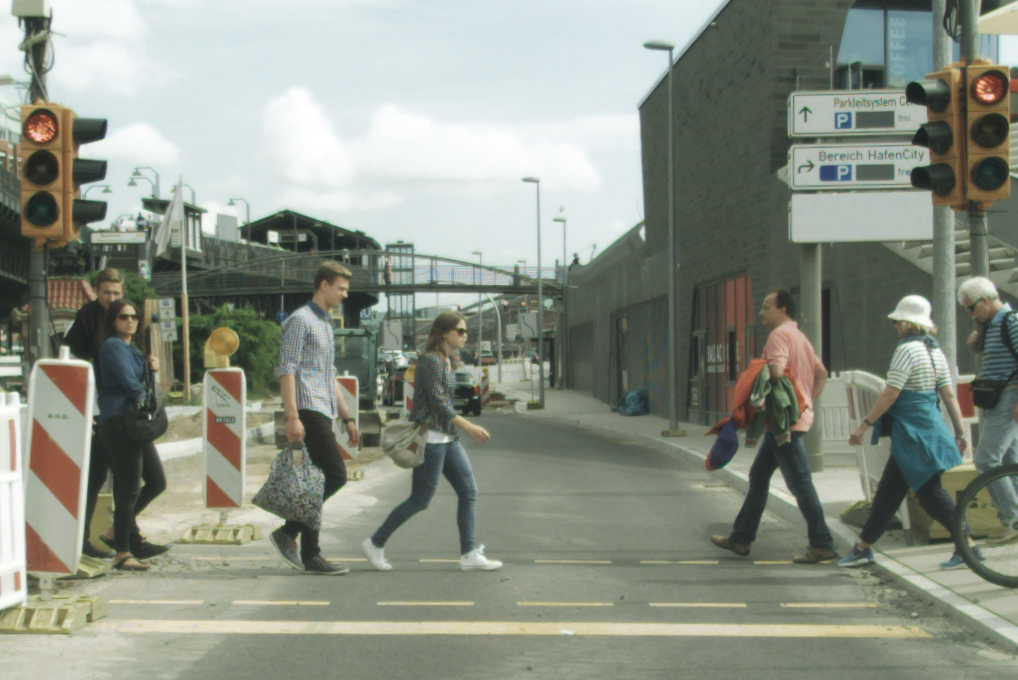} \\ \hspace{0.025in}
        }
        \subfigure{
            \includegraphics[width=1.1in]{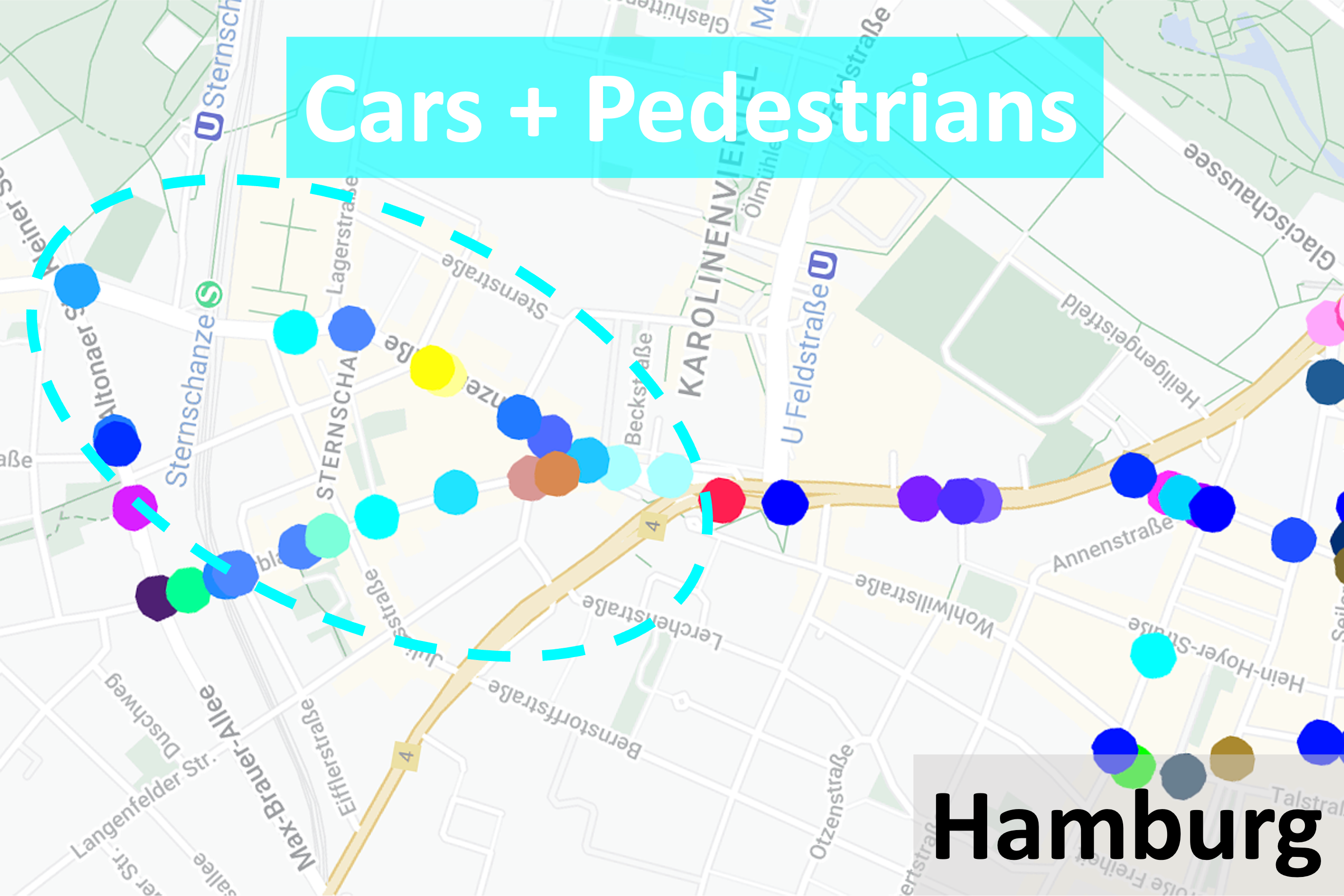} \\ \hspace{0.025in}
            \includegraphics[width=1.1in]{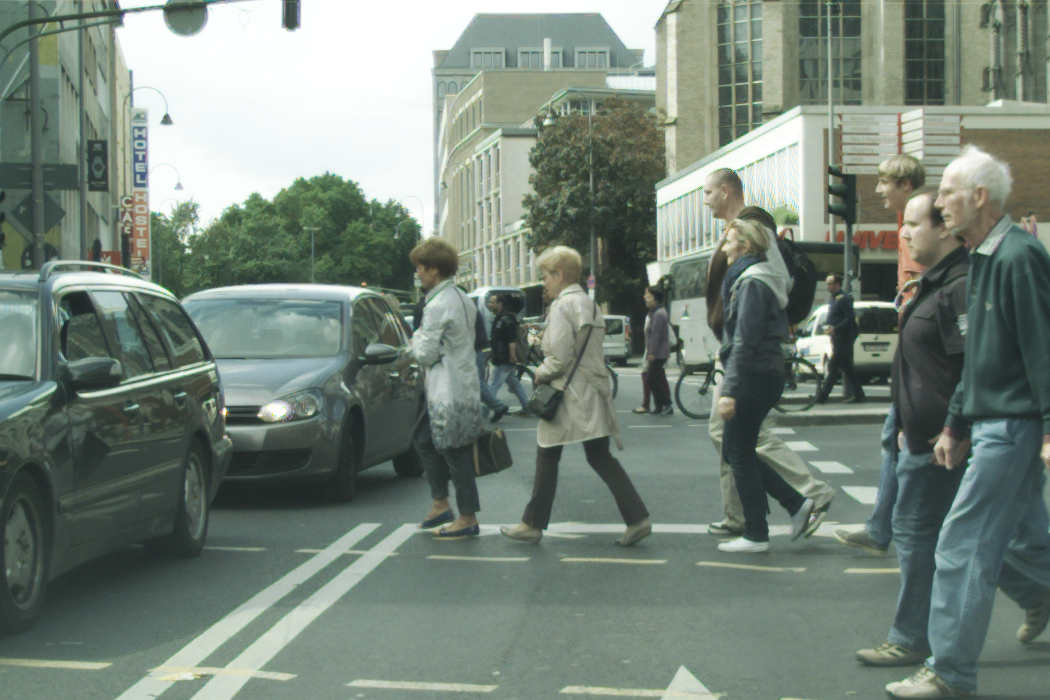} \\ \hspace{0.025in}
        }
        \subfigure{
            \includegraphics[width=1.1in]{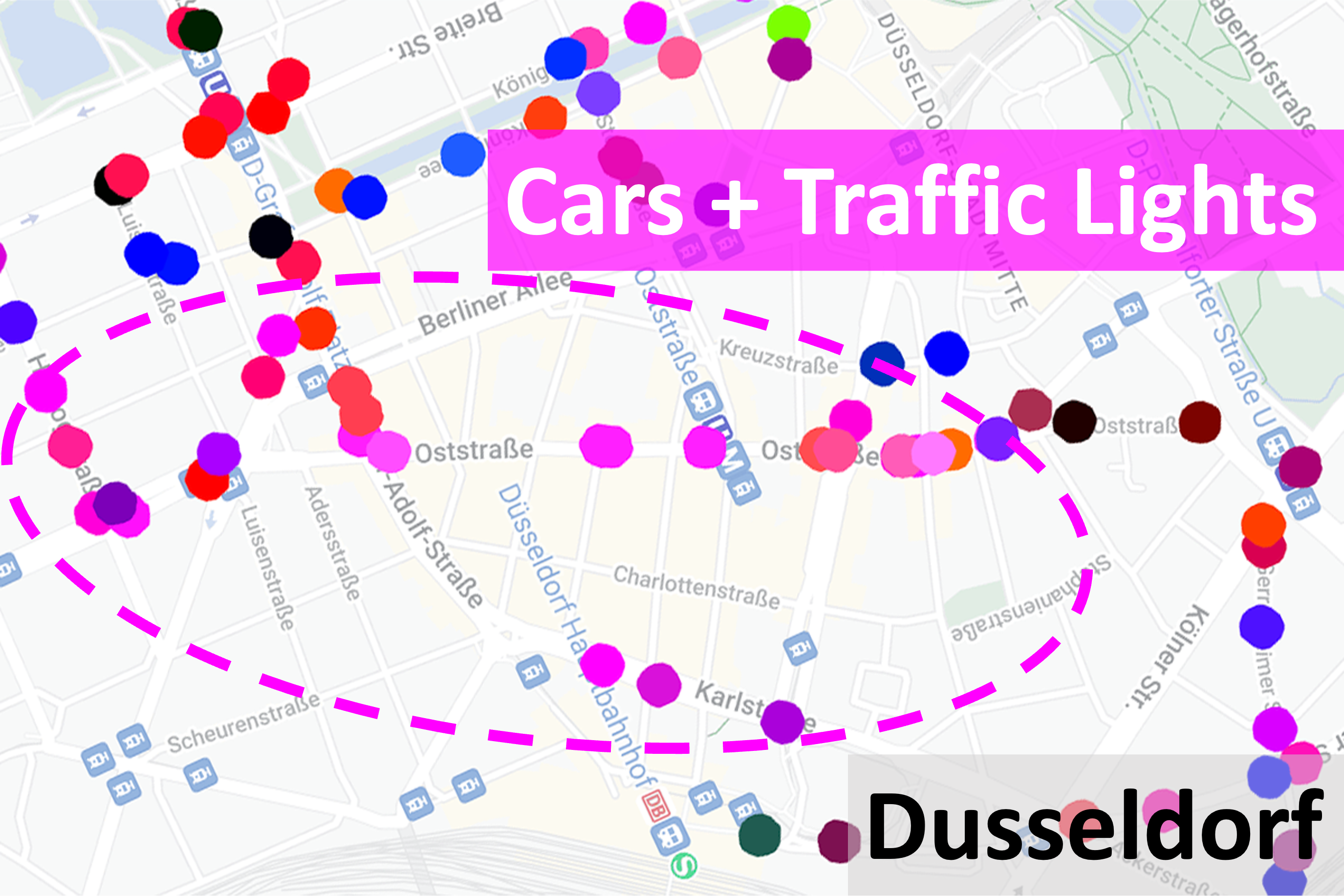} \\ \hspace{0.025in}
            \includegraphics[width=1.1in]{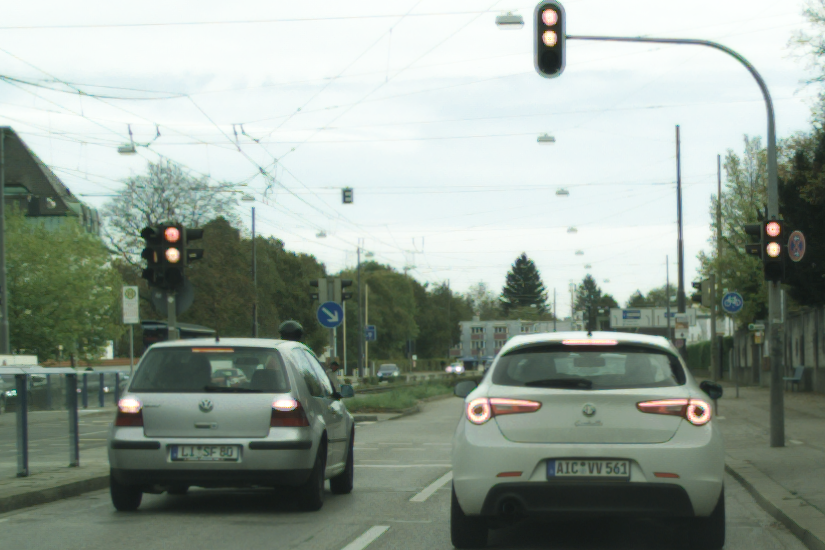} \\ \hspace{0.025in}
        }
    \end{minipage}
    \begin{minipage}[h]{0.15\linewidth}
        % \vspace{0.065in}
        \subfigure{
        % \vspace{0.065in}
            \includegraphics[width=1.1in]{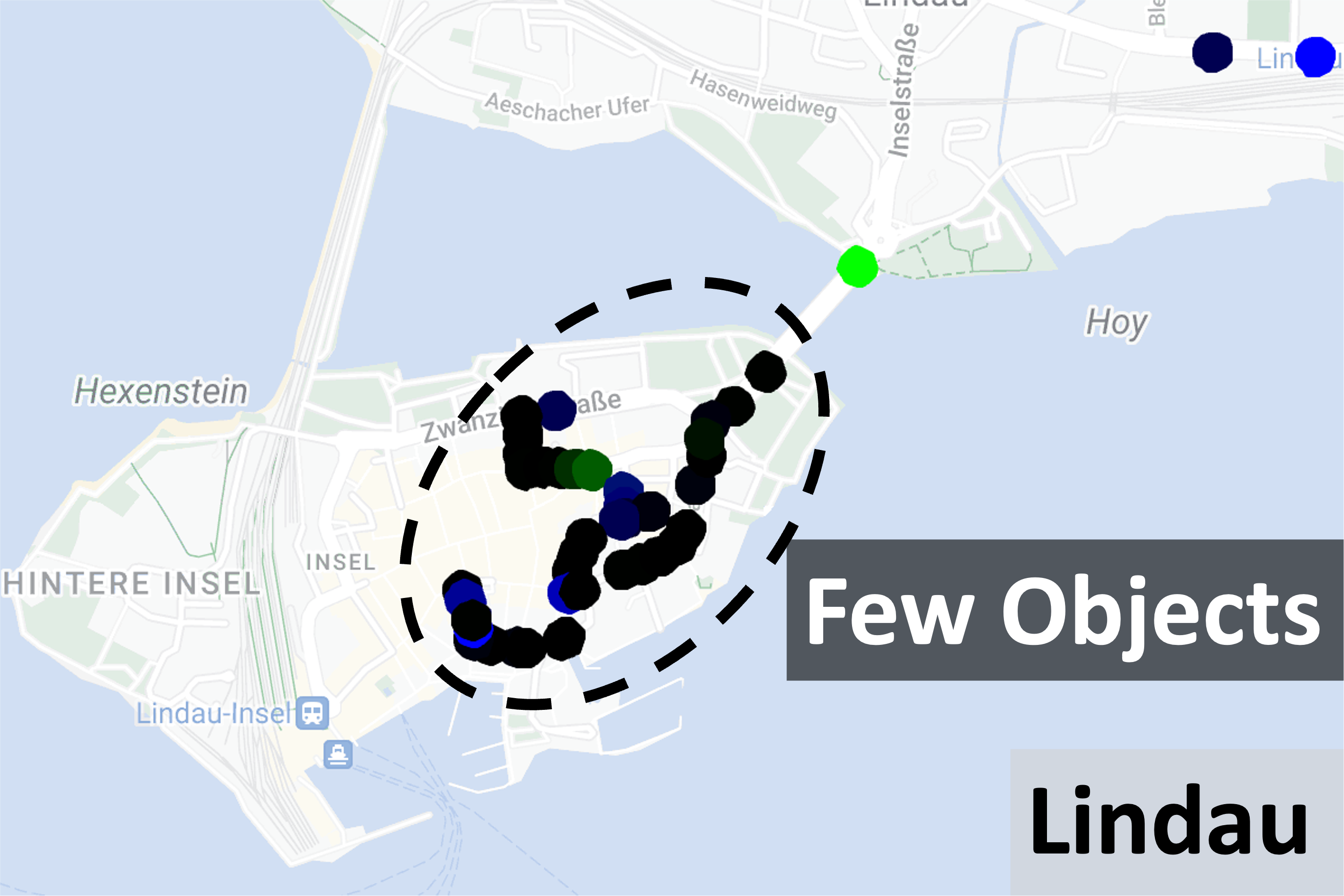} \\ \hspace{0.025in}
            \includegraphics[width=1.1in]{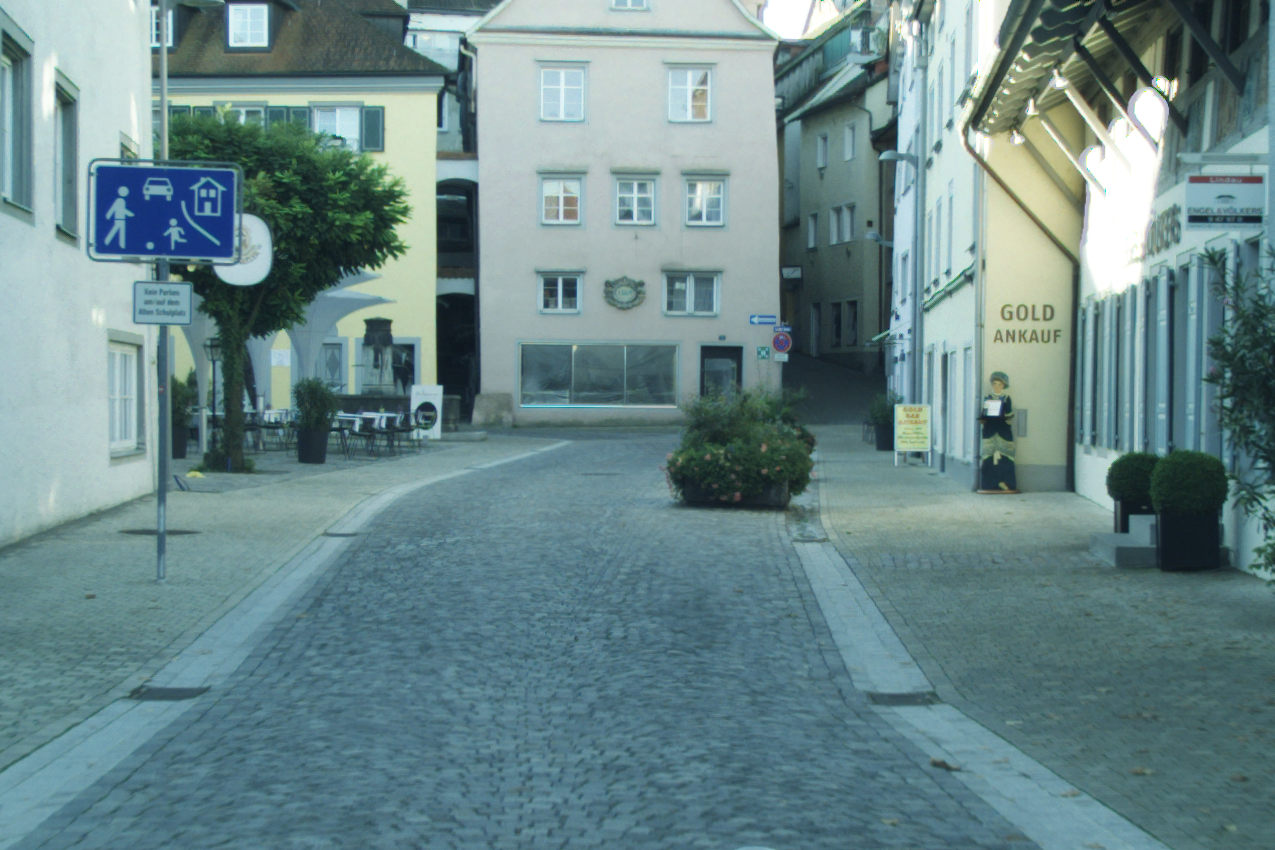} \\ \hspace{0.025in}
        }
        \subfigure{
            \includegraphics[width=1.1in]{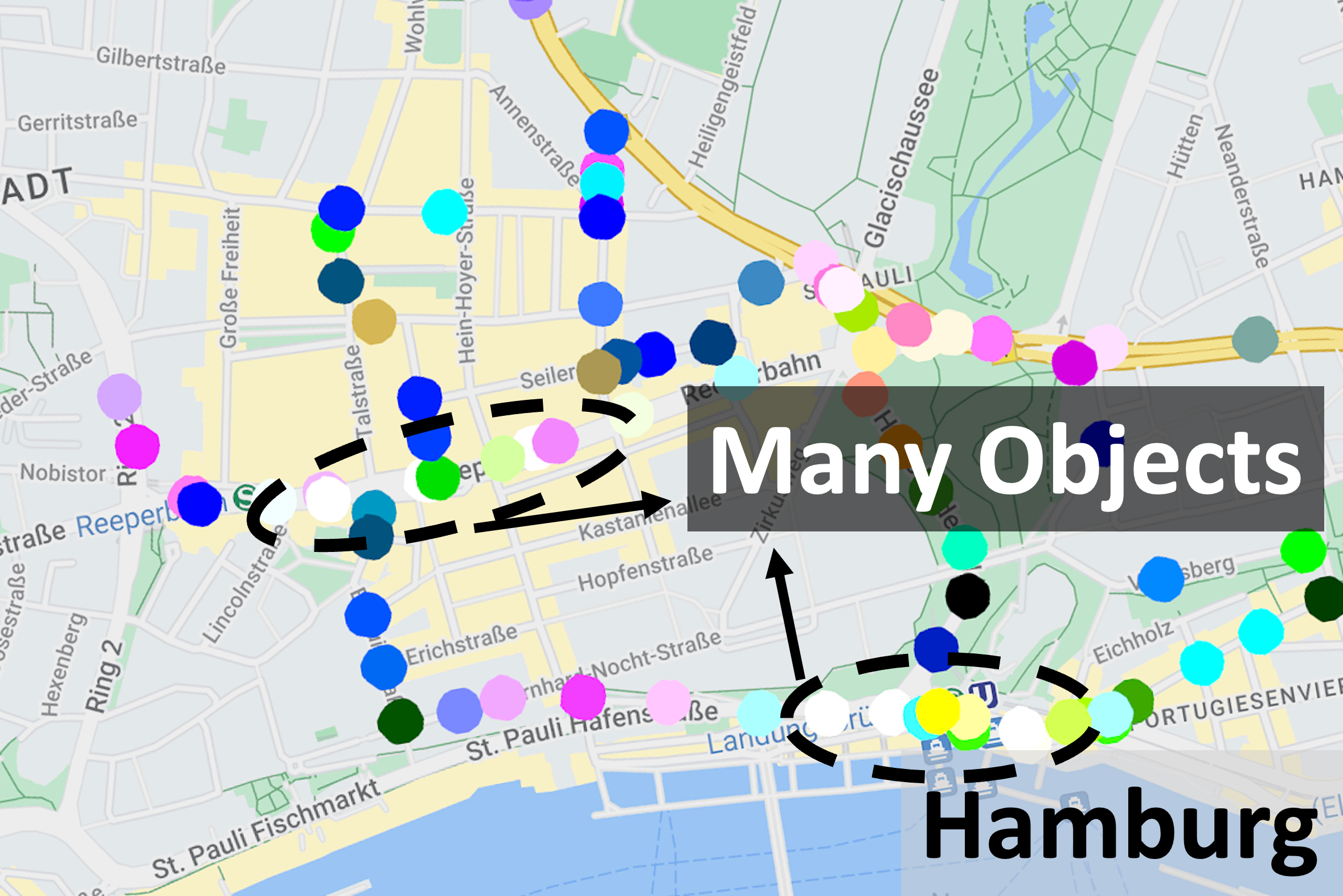}\\ \hspace{0.025in}
            \includegraphics[width=1.1in]{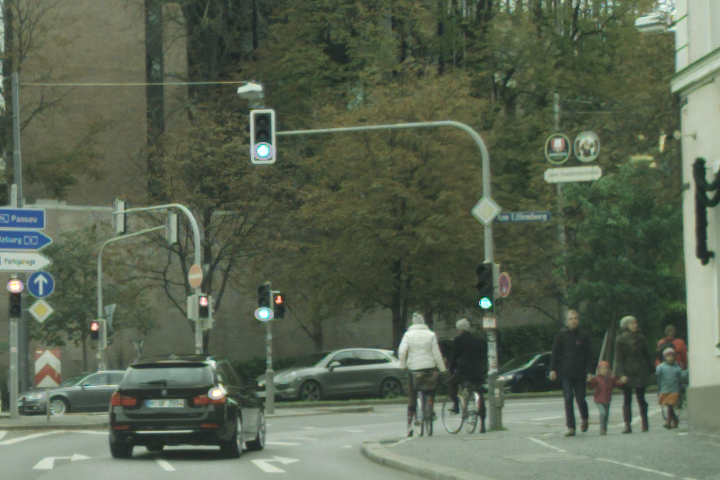} \\ \hspace{0.025in}        
        }
        \subfigure{
            {\small \hspace{1.03in}  (c)}
        }
        % \hspace{-0.001in}
        \subfigure{
            \hspace{0.005in}
            \includegraphics[width=1in]{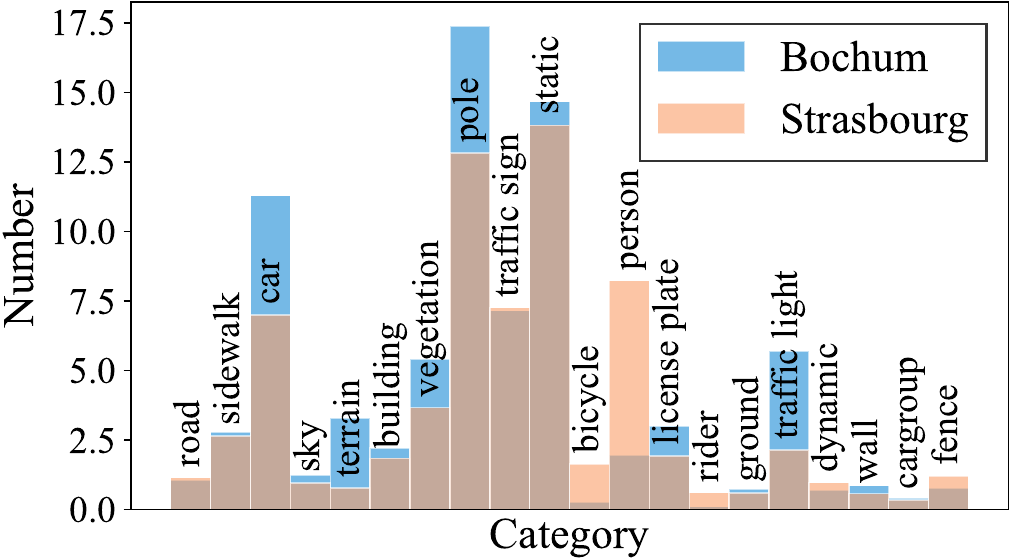}\\ \hspace{0.025in}
            \hspace{0.05in}
            \includegraphics[width=1in]{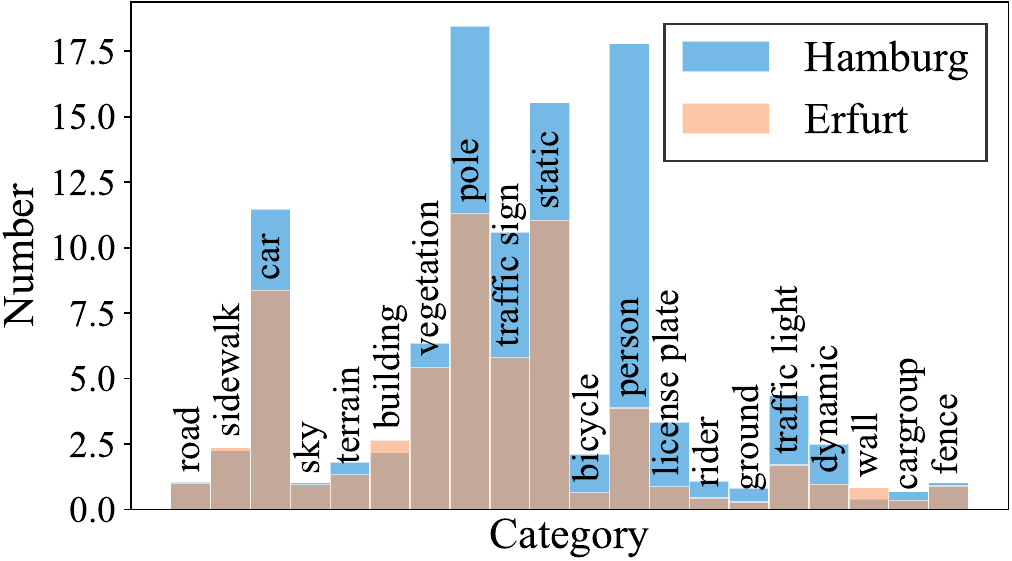}\\ \hspace{0.025in}
        }
    \end{minipage}
    \vspace{-0.15in}
}

\vspace{-0.05in}
% \vspace{0.1in} 
% \hspace{2.5in}
\centering
{\small \hspace{0.1in} (a) \hspace{2.1in} (b) \hspace{2.10in} (d) }
\vspace{-0.05in}
\caption{An example of regional similarities for the samples of Cityscapes on the map. (a) Traffic lights, pedestrians, and cars are dominant in certain-size regions, respectively. (b) All their binary combinations aggregate in certain-size regions. (c) There exist few objects or many objects in some regions. (d) The Significant difference in label distribution amongst the cities.}
\vspace{-0.15in}
% \vspace{-0.25in}
\label{data_distribution_graph1}
\end{figure*}

The remained paper are organized as following. Section \ref{sec:prelim} motivates this work with several case studies.
Section \ref{sec:partition} presents the partitioning mechanism. Section \ref{sec:FRL} shows the design of the FedRAV framework. The
experimental evaluations are presented in Section \ref{sec:experiments}. Section \ref{sec:conclusion} concludes this work.

\section{Preliminary Case Studies And Motivation}  \label{sec:prelim}
Before introducing a holistic framework, we motivate our work with a real-world example.
We claim that the spatial distribution of vehicles' collected data exhibits \textit{regional similarities}, i.e., samples with similar label distributions tend to aggregate in nearby areas, forming regions. For example, urban and rural areas are usually divided into various sub-regions according to their diverse functions, i.e., schools, stadiums, commercial centers, transportation hubs, highways, country roads, and residential roads, etc. Vehicles traversing these regions tend to encounter more similar scenarios intra-regionally and encounter less similar scenarios inter-regionally. The claim is verified via analysis of realistic data from German cities in the following.

\begin{table}[!ht]
% \small
\centering
\footnotesize
 \caption{Color mapping table.}
  \begin{tabular}{llllllll}
    \toprule
    Color & $c_i^r$ & $c_i^g$ & $c_i^b$ & Illustration of distribution  \\
    \midrule
    Red & 255 & 0 & 0 & Traffic Lights Dominant \\
    Green & 0 & 255 & 0 & Pedestrians Dominant \\
    Blue & 0 & 0 & 255 & Cars Dominant\\
    Yellow & 255 & 255 & 0 & Abundant Traffic Lights \& Pedestrians\\
    Cyan & 0 & 255 & 255 &  Abundant Cars \& Pedestrians \\
    Purple & 255 & 0 & 255 & Abundant Cars \& Traffic Lights \\
    Black & 0 & 0 & 0 & Three Categories Scarce \\
    White & 255 & 255 & 255 & Three Categories Dominant\\
    \bottomrule
  \end{tabular}
  \label{table:color_map}
\end{table}

Specifically, the Cityscapes \cite{cordts2016cityscapes}, a semantic segmentation dataset, consists of complex urban street scenes collected from $50$ cities of Germany, covering over $30$ different categories of traffic participants. The metadata of each image in the dataset provides the labels of different traffic participants and the spatial location of the collected data. We build the following visualization approach and choose three representative categories (i.e., traffic lights, pedestrians, and cars) to depict the data distribution of diverse categories. In order to intuitively depict the data distribution of samples on the map, the number of objects in each category of a sample is mapped to the color codes of RGB channels, where the colors red, green, and blue represent traffic lights $\omega_1$, pedestrians $\omega_2$, and cars $\omega_3$, respectively. The color code of RGB is an integer ranging from 0 to 255, and each color is denoted as a tuple $(c^r,c^g,c^b)$ satisfying $\{c^r,c^g,c^b\} \in [0,255]$ and $\{c^r,c^g,c^b\} \in \mathcal{N}$. More formally, the red color code $c^r_i$ of sample $i$ is defined as a floor function,
\begin{equation}
\small
c^r_i = \Big \lfloor \frac{l^{\omega_1}_i -\bar l^{\omega_1}_{\text{min}}}{\bar l^{\omega_1}_{\text{max}}-\bar l^{\omega_1}_{\text{min}}} \times 255 \Big \rfloor,
\label{eqn:red_color_code} 
\end{equation}
where $\bar l^{\omega_1}_{\text{max}} = \max\limits_{j}\{\bar l^{\omega_1}_{(j)}\}$ and $\bar l^{\omega_1}_{\text{min}} = \min\limits_{j}\{\bar l^{\omega_1}_{(j)}\}$. The subscript $j \in [M]$ denotes the $j$-th city among $M$ cities. $l^{\omega_1}_i$ represents the number of traffic lights in the sample $i$. $\bar l^{\omega_1}_{(j)}$ denotes the average quantity of traffic lights of the samples in the $j$-th city. The notations $\bar l^{\omega_1}_{\text{max}}$ and $\bar l^{\omega_1}_{\text{min}}$ represent the maximum and minimum average number of traffic lights of a sample among all $M$ cities.

Therefore, $c_i^r$ represents the richness of traffic light objects in the sample $i$. The larger value of $c_i^r$ depicts more abundant objects of traffic lights while the smaller value reflects the deficiency. $c_i^r$ is in value range $[0, 255]$ and follows the rounding rule: if $c_i^r< 0$, $c_i^r$ is rounded to $0$; if $c_i^r> 255$, $c_i^r$ is rounded to $255$. Furthermore, $c_i^g$ and $c_i^b$ are defined similarly as, $c^g_i = \lfloor (l^{\omega_2}_i -\bar l^{\omega_2}_{\text{min}})/ (\bar l^{\omega_2}_{\text{max}}-\bar l^{\omega_2}_{\text{min}}) \times 255\rfloor$ and $c^b_i = \lfloor (l^{\omega_3}_i -\bar l^{\omega_3}_{\text{min}})/ (\bar l^{\omega_3}_{\text{max}}-\bar l^{\omega_3}_{\text{min}}) \times 255\rfloor$
%$c^g_i = \frac{l^{\omega_2}_i -\bar l^{\omega_2}_{\text{min}}}{\bar l^{\omega_2}_{\text{max}}-\bar l^{\omega_2}_{\text{min}}} \times 255$ and $c^b_i = \frac{l^{\omega_3}_i -\bar l^{\omega_3}_{\text{min}}}{\bar l^{\omega_3}_{\text{max}}-\bar l^{\omega_3}_{\text{min}}} \times 255$
respectively, following the same rounding rules as $c^r_i$. $c^g_i$ and $c^b_i$ depict the richness of pedestrians and cars in sample $i$.

As shown in Fig. \ref{data_distribution_graph1}, the distributions of samples are intuitively depicted on the Google Maps of cities via the above pre-defined color codes of the RGB channel. The codes reflect the label distribution of samples.  Specifically, the Figs. \ref{data_distribution_graph1}(a)-(c) demonstrate some typical cases of the spatial distribution of samples in the Cityscapes dataset. The fundamental observation is that there exist 8 representative label distributions, whose details are summarized in Table \ref{table:color_map}. %Concretely, samples with similar label distributions tend to aggregate in approximated regions. 
It is observed that, certain patterns exist on the map, where samples with similar label distributions tend to aggregate in approximated regions.

The top two figures in Fig. \ref{data_distribution_graph1}(a) show regions where vehicles traverse multiple contiguous city blocks, where traffic lights or pedestrians are dominant in these blocks. The third figure shows a region where cars are crowded while traffic lights and pedestrians are insignificant. Fig. \ref{data_distribution_graph1}(b) demonstrates another three cases that two out of the three labels are dominant, where yellow regions show areas like harbors with many pedestrians and traffic lights, cyan regions show blocks with many cars and pedestrians appearing, and purple regions show popular commuting roads for cars together with many traffic lights but few pedestrians. 
In certain remote areas, there is scarcely any presence of traffic participants while in popular areas like tourist attractions, traffic participants of all three categories are abundant as shown in Fig. \ref{data_distribution_graph1}(c). 

%We further explore 
%The difference of label distribution in driving scenarios is further explored by analyzing 
%amongst the cities is further explored as shown in Fig. \ref{data_distribution_graph1}(d).  
The collected vehicles' data demonstrate dissimilarities across cities as shown in Fig. \ref{data_distribution_graph1}(d).
The number of objects in 20 different labels is counted for different cities, leading to the comparison between Bochum and Strasbourg, and between Hamburg and Erfurt. In the first comparison, Bochum tends to have more poles, cars, and traffic lights since it is a greater city compared with the town Strasbourg, while Strasbourg tends to be more populous since many tourists aggregate locally. In the second comparison, as one of the greatest cities in Germany, Hamburg has more cars, poles, traffic signs and persons compared with Erfurt. Therefore, factors such as population, geography, culture, and economic development all affect the exhibited label distributions of samples collected in different cities, which substantiates that the spatial distributions of collected vehicle data do vary across regions. Based on the analysis above, we summarize the motivation below.

\textbf{Motivation} \textit{The spatial distribution of vehicles' collected data exhibits visible regional similarities. 
% Specifically, 
The data distributions are more similar intra-regionally but less analogous inter-regionally. Therefore, partitioning regions adaptively is necessary for realizing FL in AVs.}
%Therefore, realizing FL in AVs and considering partitioning regions adaptively is necessary.}

\section{Regional Partitioning}  \label{sec:partition}
Motivated by the regional similarities of the spatial distribution of AVs' collected data, we introduce a novel %Euclidean distance 
distance in metric space to describe the similarity properties, which is referred to as \textit{Region-Wise Distance(RWD)}.
Furthermore, we propose a new partitioning algorithm to divide large areas into sub-regions.

\subsection{Problem Formulation}

We consider a vehicular network with $N$ AVs (a.k.a. clients)
% \footnote{We use AV, client, and user interchangeably, henceforth. }
, $K$ region servers, and one central server. AVs distributed across $M$ cities or towns can traverse several city blocks along the 
% grid-like 
roads. 
% [zz]%需要强调grid-like吗？现实道路不行？
Each AV $i \in [N]$ has a local dataset $\mathcal{D}_i = \{x_{(i,j)}, y_{(i,j)}\}^{\vert \mathcal{D}_i \vert}_{j=1}$ where $x_{(i,j)}$ is trainable image and $y_{(i,j)}$ is true label for the image classification task.
% [zz]%x还有y是什么？你没解释。我这里看不太明白这个D_i里面到底是什么数据，第i辆车上面采集的所有数据，比如一段时间里的连续1000张图片？那为什么下面V_i部分又只有一个坐标，1000张图片的坐标不应该不同吗？
% with $\vert \mathcal{D}_i \vert$ samples drawn from its underlying distribution $ \mathcal{\hat D}_i$
% and records its own real-time vehicular coordinate $V_i = [v_x, v_y]$. 
$V_i = [v_x, v_y]$ represents the coordinates of AV $i$.
$L_i=[l^1_i, l^2_i,l^3_i,\cdots, l^m_i]$ is the label vector of class distribution of $\mathcal{D}_i$ for $i$-th AV, where $l^m_i$ is the number of objects in $m$-th semantic category for AV $i$.  We assume that a large area containing $N$ AVs is divided into $K$ sub-regions, i.e., $ \{ \mathcal{A}_k \}^{K}_{k=1}$ , and $U$ is the set of $k$ regional centroids. Let $\mathcal{X} = \{(V_i, L_i)\}^N_{i=1} $ denote a set of $N$ AVs. We formulate the problem of finding optimal regional structure $\{ \mathcal{A}_k \}^{K}_{k=1}$  and regional centroids $U$ as \textbf{Regional Structure Optimization} (RSO) problem.  The quantization error $\phi_{U}(\mathcal{X})$ is minimized, where
\begin{equation}
\small
    \textbf{P1}: \ \ \ \ \ \phi_{U}(\mathcal{X}) = \sum_{x \in \mathcal{X}} \min_{u \in U} \textit{RWD}\left(x, u\right)^2
\end{equation}
\textbf{ Subject to }
\begin{equation}
\begin{array}{lc}
\small
    \textit{RWD}\left(x, u\right)=|| V_x -  V_u ||_2 +  \gamma \cdot [\zeta (C_x -  C_u)^T W \zeta (C_x -  C_u)]^{\frac{1}{2}} \\
\end{array}
\end{equation}
\begin{equation}
\begin{array}{lc}
\small
    |U|= K, \vert \mathcal{X} \vert = N & \\
\end{array}
\end{equation}
The notation $\zeta( \cdot )$ represents the element-wise absolute value of the inside vector. The first constraint represents region-wise distance between AV $x$ and regional centroid $u$ where $\gamma$ is a hyperparameter controlling importance of the two components, $W \in \mathbb{R}^{m \times m}$ is a weight matrix and $C_x$ is $M$-relative abundance vector of AV $x$. Region-wise distance measures the similarity of data distributions for AVs in geographic space and embedding space of abundance. The abundance vector $C_x$ defines the relative richness of the categories in the local data of AV $x$.
 % [zz]%一句话总结你定义的RWD的物理含义
 Specifically, we will discuss the details of the  concepts of abundance and region-wise distance in Section \ref{subsec:rwd}. The second constraint limits the number of regions and AVs. Now, the optimization problem aims to find $K$ subsets of $N$ AVs such that each AV can be assigned to a proper sub-region and data distributions are more similar intra-regionally. Thus, AVs in the same region $\mathcal{A}_k$ form a federation to jointly complete downstream learning tasks. 
 % For simplicity, the optimal quantization error is denoted by $\phi_{OPT}(\mathcal{X})$. 
 $\phi_{U}(\mathcal{A}_k)$ denotes the contribution of $\mathcal{A}_k \subset \mathcal{X}$ to the error. Then,  $\delta(\mathcal{A}_k)$ is used as an operator to indicate the mean coordinate of AVs indexed by $\mathcal{A}_k$, i.e., $\delta(\mathcal{A}_k)=\frac{1}{\vert\mathcal{A}_k\vert}\sum_{i \in \mathcal{A}_k}V_i$  and  use $\tau(\mathcal{A}_k)$ to indicate the mean abundance of AVs indexed by $\mathcal{A}_k$, i.e., $\tau(\mathcal{A}_k)=\frac{1}{\vert \mathcal{A}_k \vert}\sum_{i \in \mathcal{A}_k}C_i$. 
 % The main notations are shown in Table \ref{table:notations}.
 % [zz]%你用的符号太多了，你在第三章需要画一张table of important notations的表，不然审稿人记不得你这么多符号什么意思
 %你这里abundance的概念定义的不清楚，审稿人搞不明白什么意思

% \begin{table}[!ht]
% \centering
% \small
%  \caption{\small Summary of main notations.}
%   \begin{tabular}{ll}
%     \toprule
%     Notation & Description  \\
%     \midrule
%     % $\mathcal{D}_i$  & Local dataset of AV $i$ \\
%     $V_i$ & Vehicular coordinate of AV $i$ \\
%     $L_i$ & Feature vector of class distribution of AV $i$ \\
%     $C_i$ & M-relative abundance vector of AV $i$ \\
%     $\mathcal{X}$ & The set of $N$ AVs \\
%     $\mathcal{A}_k$ & The set of AVs in region $k$ \\
%     $U$ & The set of regional centroids \\ 
%     $\boldsymbol{w}$ & Model parameter \\
%     $\Delta \boldsymbol{w}$ & Model update \\
%     $F_i(\cdot;\boldsymbol{w}_i)$ & The local objective function of AV $i$ \\
%     $f_k(\cdot;\boldsymbol{w}_k)$ & The regional objective function of region $k$ \\
%     $\boldsymbol{v}$ & Embedding vector of hypernetwork \\
%     $\boldsymbol{\varphi}$ & Model parameter of hypernetwork \\
%     \bottomrule
%   \end{tabular}
%   \label{table:notations}
%   \vspace{-0.1in}
% \end{table}

\subsection{Region-Wise Distance} \label{subsec:rwd}
In a conventional partitioning setting, the large area is divided into 
% connected 
sub-regions based on the $\ell_2$-norm of spatial locations of AVs. However, it does not consider similarities in label distribution across AVs. 
% One arisen problem is how to maximize the \textbf{intra-regional label distribution similarities} while minimizing \textbf{inter-regional spatial fragments} 
% One arisen problem is 
% in a federated regional partitioning setting. 
% [zz]%什么是spatial fragments，看不明白
There is a trade-off between spatial location and label distribution in realistic situations. We tackle this problem via a control knob to combine \textit{spatial location} with \textit{label distribution}, which is described as follows.

\noindent \textbf{Definition 1 (Spatial Distance).} Assumed that $N$ AVs are located in the same Euclidean plane, and $V_i = [v_x, v_y]$ is the coordinate of AV $i$. Then, the spatial distance between AV $i$ and AV $j$ is defined by,
\begin{equation}
\small
    d_{\text{spatial}}(i,j) = \Vert V_i - V_j \Vert_2,  \  \forall i,j \in [N].
\end{equation}

Existing studies have used the cross-class $\ell_2$-norm distance to evaluate the similarity between label distributions. However, these measures are sensitive to changes in the number of objects of a particular class. Due to the absolute scale of these measures, regional partitioning may not benefit from them in realistic and challenging heterogeneous data distribution. Inspired by the visualization method in Section \ref{sec:prelim}, we extend the case to an M-dimensional label vector.

\begin{figure}
    \centering
   \includegraphics[width=3in]{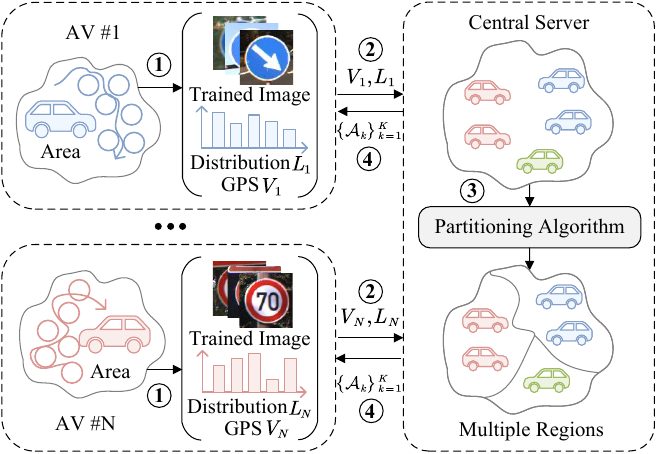}
    \caption{Illustration of partitioning mechanism. \textcircled{1} Collect data; \textcircled{2} upload coordinate $V$ and feature vector $L$; \textcircled{3} divide areas into regions; \textcircled{4} download regional structure $\{\mathcal{A}_k\}^K_{k=1}$.}
    \label{fig:tire-one}
    \vspace{-0.2in}
\end{figure}

\noindent \textbf{Definition 2 ($M$-Relative Abundance).} Given the label vector  $L_i \in \mathbb{R}^m$ of local dataset $\mathcal{D}_i$ of AV $i$, the M-Relative Abundance of AV $i$ amongst all the AVs cross the $M$ cities is represented as $C_i=[c^1_i,c^2_i,c^3_i \cdots,c^{m}_i]$. Concretely, the abundance of $m$-th category $c^m_i$ is defined as,
\begin{equation}
\small
c^m_i= \Big \lfloor \frac{l_i^{m}-\bar{l}_{\text {min }}^{m}}{\bar{l}_{\text {max }}^{m}-\bar{l}_{\text {min }}^{m}} \times 255 \Big \rfloor,
\label{eqn:color_code_m} 
\end{equation}
where we let $\bar l^{m}_{\text{max}} = \max\limits_{j}\{\bar l^{m}_{(j)}\}$ and $\bar l^{m}_{\text{min}} = \min\limits_{j}\{\bar l^{m}_{(j)}\}$. The subscript $j \in [M]$ denotes the $j$-th city among $M$ cities. $l^{m}_i$ represents the number of $m$-th category in the AV $i$. 
$\bar l^{m}_{(j)}$ denotes the average quantity of $m$-th category of all AVs in the $j$-th city.
The notations $\bar{l}^m_{\text {max}}$ and $\bar{l}^m_{\text {min}}$ represent the maximum and minimum average number of  $m$-th category of a AV among all $M$ cities, respectively. Specifically, Eq.(\ref{eqn:color_code_m}). follows the same rounding rules as Eq.(\ref{eqn:red_color_code}). 
% Ranging from 0\% to 100\%
Ranging from 0 to 255, $c^m_i$ demonstrates the relative abundance of category $m$ for AV $i$.  
%We also call $C_i$ as a special color just as $C_i$ is a color tuple of RGB channel satisfying $m = 3$. 
% $C_i$ may be explained as a special color in high-dimensional space, since $C_i$ is a color tuple of typical RGB channel if $m = 3$. 

\noindent \textbf{Definition 3 (Label Distance).} Meanwhile, the label distance is defined by,
\begin{equation}
\small
    d_{\text{label}}(i,j) = \Vert C_i - C_j \Vert_2, \  \forall i,j \in [N].
\end{equation}
It presents a geometrical evaluation in embedding space of $M$-relative abundance vector. The generic form of the distance is given as follows,
\begin{equation}
\small
    d_{\text{label}}(i,j)= [\zeta (C_i -  C_j)^T W \zeta (C_i -  C_j)]^{\frac{1}{2}}, \ \forall i,j \in [N],
    \label{eqn:weighted_color_distance}
\end{equation}
where $W \in \mathbb{R}^{m \times m}$ is a weight matrix satisfying 
% $\sum_{i=1}^m\sum_{j=1}^mW_{i j}=1$ and 
$W_{i j} \geq 0$. Here, we let $W=I$, which means that all categories are equally essential.  
% [zz]%W所有元素求和等于1？W=I的时候，所有元素求和不是m吗？写错了吧？

In order to obtain a optimal regional structure satisfying data similarities, 
% and connectedness
we propose to apply region-wise distance to the optimization problem \textbf{P1}. As discussed earlier, a key challenge here is %how to trade them off.
to achieve trade-off between spatial locations and label distribution.
Therefore, a control knob  $\gamma \in [0, 1]$  is proposed to balance the spatial distance and label distance.

\noindent \textbf{Definition 4 (Region-Wise Distance).} Region-Wise Distance between AV $i$ and AV $j$ is defined by,
\begin{equation}
\small
    \textit{RWD}(i,j) = || V_i -  V_j ||_2 +\gamma \cdot [\zeta (C_i -  C_j)^T W \zeta (C_i -  C_j)]^{\frac{1}{2}},
\end{equation}
where $\forall i,j \in [N]$ and $\gamma$ is adjustable hyperparameter. 
% Region-wise distance measures the similarity of data distributions for AVs in geographic space and embedding space of abundance while reducing fragments. 
% [zz]%你这里的fragments包括connectedness都介绍得不清楚，难以清除理解含义

\subsection{Partitioning Mechanism}

Our partitioning mechanism follows a client-server paradigm with one-shot communication, as shown in Fig. \ref{fig:tire-one}. 
% [zz]%AV和AEV要统一，你这篇文章叫的是AV，所以图2里面包括其它地方的AEV全部都要替换成AV。
%图2里面应该是Trained Image吧？
The mechanism's crux is thoroughly capturing the similarity between AVs by computing the region-wise distance. Therefore, multiple AVs are required to send their coordination $V$ and label vector $L$ to a central server. 
% For privacy security, they can transmit encrypted quantities of the objects in an order specified by the server instead of transmitting privacy-sensitive information about category names. 
% [zz]%你这句是想说保护隐私吗？vector每个位置对应的category name是固定的，传每个位置的值，要知道category name很容易吧
% This mechanism is compatible with various existing encryption approaches, e.g., homomorphic encryption \cite{mahato2023comparative}. 
On the server side, we execute a partitioning algorithm to divide large areas of interest into sub-regions and send regional structure $\{\mathcal{A}_k\}^K_{k=1}$ back to AVs.
% [zz]%检查A是否应该有下标k

Specifically, the \textbf{P1} is analogous to the \textit{Euclidean Clustering Problem}, which is unfortunately NP-hard even for $K=2$. Due to NP-hardness, no polynomial-time algorithm exists unless P equals NP. Lloyd proposed a simple and fast heuristic called Lloyd's algorithm\cite{lloyd1982least} that begins with $K$ arbitrary initial solution and iteratively converges to a locally optimal solution. However, Lloyd's algorithm is susceptible to improper initialization. K-Means++ seeding\cite{arthur2007k} can provide provably good initialization with $8(lnK+2)$-approximation to the global optimum in expectation. Inspired by the prior works, we propose a novel partition algorithm 
% with an initial approximation guarantee 
to solve \textbf{P1} as shown in Algorithm \ref{alg: partition_algorithm}. The algorithm first samples an initial centroid uniformly at random from the set of AVs $\mathcal{X}$ and adaptively samples $K - 1$ additional centroids using seeding steps (lines 1-4). In each seeding iteration, the AV $x \in \mathcal{X}$ is added to the set of already sampled centroids $U$ with probability $\frac{\phi_{U}(\{x\})}{\phi_{U}(\mathcal{X})}$ where
% $\phi_{U}(\{x\})$ is the contribution of $\{x\}$ to error 
$\phi_{U}(\{x\})$ is the shortest region-wise distance from AV $x$ to the closest centroid that the Algorithm \ref{alg: partition_algorithm} has chosen.
% and $\phi_{U}(\mathcal{X})$ is quantization error.
% We prove that the initial solution satisfies the $O(logK)$ approximation bound, as shown in Theorem 1.
Then, the algorithm will use region-wise distance as a metric to repeat the standard Lloyd's steps (lines 5-8)  until a proper regional structure is found.

\begin{algorithm}[t!]
\caption{Regional Partitioning Algorithm}
\label{alg: partition_algorithm}
% \small
\footnotesize

\KwIn{ A set $\mathcal{X}$ of $N$ AVs,  number of regions $K > 0$,  control knob $0 \leq \gamma \leq 1$, weight matrix $W \in \mathbb{R}^{m \times m}$.}

\KwOut{regional structure $\{ \mathcal{A}_k \}^{K}_{k=1}$ and regional centroids $U$.}

$U \gets$ Sample an AV uniformly at random from $\mathcal{X}$

    \While{$\vert U \vert < K$} 
    {
        Sample $x \in \mathcal{X}$ with probability $\frac{\phi_{U}(\{x\})}{\phi_{U}(\mathcal{X})}$  \Comment{Sample centroids}
            
        $U \gets U \cup \{x\}$ 
    }

    \While{$U$ is not stable} 
    {
        $\mathcal{A}_r \gets \{i: \textit{RWD}(i, r)^2 \leq \textit{RWD}(i, s)^2, \forall s \in U \}$ 
        
        $V_r = \delta(\mathcal{A}_r)$ , $C_r=\tau(\mathcal{A}_r)$  

        Update $U$ by $(V_r, C_r)$ \Comment{Update centroids}
    }

    \textbf{return}  $\{ \mathcal{A}_k \}^{K}_{k=1}$ and $U$

\end{algorithm}
%\vspace{-0.15in}

\begin{figure*}[!ht]
     \centering
    \includegraphics[width=6.8in]{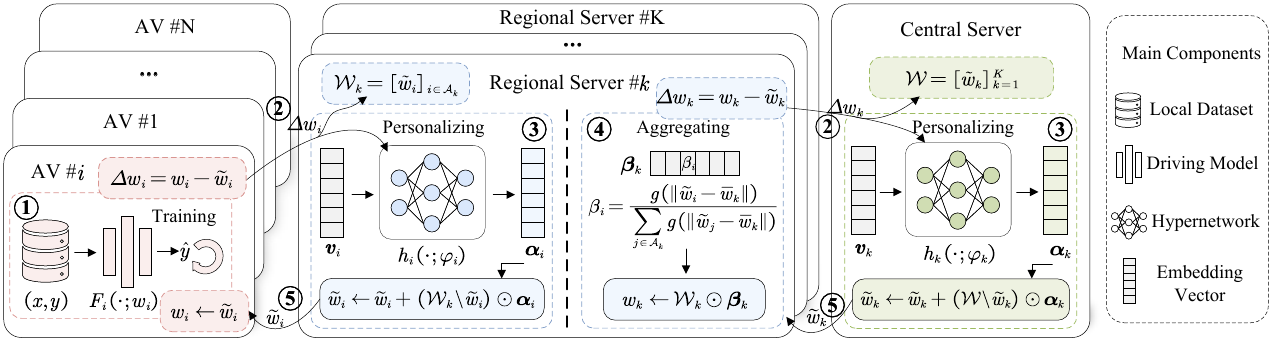}
     \caption{An overview of FedRAV Framework. The main operations consist of: \textcircled{1} local training on private dataset $\mathcal{D}$ ; \textcircled{2} upload model update $\Delta \boldsymbol{w}$; \textcircled{3} model personalization; \textcircled{4} intra-region model aggregation; \textcircled{5} download personalized model $\boldsymbol{\tilde{w}}$.}
     \label{fig:tire-two}
     \vspace{-0.15in}
\end{figure*}

\section{Federated Region-learning Framework of Autonomous Vehicles (FedRAV)}  \label{sec:FRL}

In this section, we present 
% our proposed 
the FedRAV framework that learns driving models via device-edge-cloud communication.
% The FedRAV uses hypernetworks to achieve that the personalized vehicular model adopts the beneficial models while discarding the unprofitable ones.

\subsection{Problem Formulation}
In the standard FL setting, the optimization objective aims to find a single global model. It assumes the global model is capable of learning empirical knowledge for every case. However, the prior studies like \cite{zhao2018federated} have proved that a single model may not fit parameter space from models of multiple users. The leading cause of the problem is that local data heterogeneity leads to 
% heterogeneous 
discrepant model parameter space.
% [zz]%heterogeneous space的说法有些怪
More specifically, the discrepant model disturbs the inter-client knowledge transfer. Therefore, to resolve these problems, we propose \textbf{FedRAV framework} that learns a personalized  model for each AV and each region respectively. 
% The optimization objective of FedRAV is described as follows.

The optimal regional structure $\{ \mathcal{A}_k \}^K_{k=1}$ has been obtained in Section \ref{sec:partition}. AV $i$ expects to learn personalized vehicular models $\boldsymbol{w}_i$  
% \footnote{The universal symbol $\boldsymbol{w}$ is driving model. The $\boldsymbol{w}_i$ and $\boldsymbol{\tilde{w}}_i$ represent the personalized vehicular model for AV $i$, where $\boldsymbol{w}_i$ is on the client side and $\boldsymbol{\tilde{w}}_i$ is on the regional server side. Similarly, $\boldsymbol{w}_k$ and $\boldsymbol{\tilde{w}}_k$ represent the personalized regional model for region $k$, where $\boldsymbol{w}_k$ is on the regional server side and $\boldsymbol{\tilde{w}}_k$ is on the central server side.} 
% [zz]%你这个footnote里的东西写法是不规范的，w_i还有w_k人们会认为是一个东西，只是下标不同，结果你这里表示三种不同的model。你得强调，这是一个通用符号，用到边-端，端-云，都行。两个w，表示一对局部与更广的全局的关系。
on private dataset $\mathcal{D}_i$. Thus, the local objective $F_i(\cdot;\cdot)$ of $i$-th vehicle is defined as,
\begin{equation}
\small
    F_i(\mathcal{D}_i;\boldsymbol{w}_i) \triangleq \frac{1}{\vert \mathcal{D}_i \vert} \sum_{j=1}^{\vert \mathcal{D}_i \vert} \ell\left( x_{(i,j)}, y_{(i,j)};\boldsymbol{w}_i \right),
\label{eqn:local_empirical_loss}
\end{equation}
where  $\ell(\cdot ; \cdot)$ represents a user-defined loss function. In the training of each region $k$ in FedRAV, the vehicle-level joint optimization objectives are formulated as, 
\begin{equation}
\small
    \textbf{P2}: \ \ \ \ \ \mathcal{W}_{k}^{*}=\argmin_{\mathcal{W}_{k}}
    \frac{1}{\left| \mathcal{A}_k \right|}\sum_{i\in \mathcal{A}_k}{F_i}\left( \mathcal{D}_i;\boldsymbol{w}_i\right),
    \label{eqn:p2}
\end{equation}
where optimization variable $\mathcal{W}_k=[\boldsymbol{w}_i]_{i \in \mathcal{A}_k}$ represents the set of personalized vehicular models within the region $\mathcal{A}_k$. The optimization problem aims to find a set of personalized vehicular models fitting AVs' parameter space, respectively.
From a regional perspective, region $k$ also expects to learn personalized regional models $\boldsymbol{w}_k$ on union dataset $\mathcal{D}_k=\underset{i\in \mathcal{A}_k}{\cup}\mathcal{D}_i$. Specifically, AVs need not to transmit their own privacy-sensitive datasets to each region server or central server. Therefore, the regional objective $f_k$ is defined by,
\begin{equation}
\small
f_k\left(\mathcal{D}_k; \boldsymbol{w}_k \right) \triangleq \frac{1}{\left| \mathcal{D}_k \right|} \sum_{j=1}^{\vert \mathcal{D}_k \vert} \ell\left( x_{(k,j)}, y_{(k,j)};\boldsymbol{w}_k \right).
\end{equation}
We 
% also
formulate region-level joint optimization objectives as,
\begin{equation}
\small
    \textbf{P3}: \ \ \ \ \ \mathcal{W}^{*}=\argmin_{\mathcal{W}}\frac{1}{K}\sum^K_{k=1}{f_k}\left( \mathcal{D}_k;\boldsymbol{w}_k\right),
    \label{eqn:p3}
\end{equation}
where the optimization variable $\mathcal{W}=[ \boldsymbol{w}_k ]^K_{k=1}$ is the set of all the personalized regional models. Similarly, the optimization problem seeks to find a set of personalized regional models fitting parameter space related to regions.

\subsection{Overview of FedRAV Framework}

We now present our FedRAV framework that can resolve the above optimization problems \textbf{P2} and \textbf{P3} via hypernetworks. Specifically, hypernetworks are widely used in fields of natural language modeling and computer vision to generate model parameters for other neural network.
% We apply hypernetworks for each AV and region on the server side to generate personalized mask vectors. 
Before delving into more detail, the overview of the federated optimization process is shown in Fig. \ref{fig:tire-two}. Concretely, the entire optimization process consists of the following three stages.
% : 1) the federated initialization stage; 2) the local personalization stage and 3) the regional personalization stage.

\textbf{Federated Initialization.} As shown in Fig. \ref{fig:tire-two}, each AV has a designated hypernetwork on the regional server side, and each region holds a corresponding hypernetwork on the central server side. Then, the hypernetwork $h(\boldsymbol{v}; \boldsymbol{\varphi})$ is also a particular deep neural network that maintains a trainable embedding vector $\boldsymbol{v}$ as input and model parameter $\boldsymbol{\varphi}$. Before the FedRAV starts training, the embedding vector $\boldsymbol{v}$ and model parameter $\boldsymbol{\varphi}$ are initialized randomly. Given $\boldsymbol{v}$, the hypernetwork $h(\cdot;\cdot)$ can output the personalized mask vectors $\boldsymbol{\alpha}$ which can learn cross-client or cross-region information.

\textbf{Local Personalization.} Without loss of generality, suppose that AV $i$ executes local training with $\kappa_1$ iterations on the private dataset $\mathcal{D}_i$ and sends model update $\Delta \boldsymbol{w}_i$ to regional server $k$, as shown in the left part of Fig. \ref{fig:tire-two}. Then, the regional server $k$ can update the embedding vector $\boldsymbol{v}_i$ and model parameter $\boldsymbol{\varphi}_i$ via model update $\Delta \boldsymbol{w}_i$. Specifically, we will discuss the details of the update process in the next Section \ref{sec:personalization}. As shown in the middle part of Fig. \ref{fig:tire-two}, the local personalization process of FedRAV takes place on the regional server $k$. Based on the updated embedding vector $\boldsymbol{v}_i$, the hypernetwork $h_i(\cdot;\cdot)$ can generate the new mask vectors $\boldsymbol{\alpha}_i$. FedRAV can obtain personalized vehicular models by using the mask vectors to
weight the models shared by vehicles within the same region.
% discard some unprofitable models and 
%weight
% [zz]%这里我不知道哪个词更合适，感觉weight不对，你看一看
% keep
% other beneficial models. 
Unlike FL approaches 
aggregate
%parameterized by 
a single global model, FedRAV may fit each AV's model parameter space, which gets rid of
%alleviates the interference of 
unprofitable models and ensures a better knowledge transfer.

\textbf{Regional Personalization.} We utilize the same trick of local personalization to personalize the regional model. Also, FedRAV provides a heuristic aggregation policy to obtain the regional model update $\Delta \boldsymbol{w}_{k}$, which is used to update the embedding vector $\boldsymbol{v}_{k}$ and hypernetwork model parameter $\boldsymbol{\varphi}_{k}$ in central server side, as shown in the right part of Fig. \ref{fig:tire-two}. We will discuss our heuristic policy for the intra-regional aggregation in Section \ref{sec:aggregation}. 
% Compared with CFL-based algorithm \cite{ghosh2020efficient, sattler2020clustered, vahidian2023efficient}, FedRAV framework decouples the size of the hypernetwork with the communication process, which does not induce additional communication burden.
% The regional personalization phase is not executed for every communication round, which depends on the downstream-specific learning task. In contrast, regional servers and central server indirectly undertake the computational and storage overheads of the model personalization. In large-scale vehicle federation learning settings, the 
%layered property 
% heterogeneity of FedRAV fulfills the learning tasks' needs in different layers while alleviating communication congestion.   
The details are summarized in Algorithm \ref{alg:frl}.
% The training procedures are executed alternately and repeated until convergence.

\begin{algorithm}[t!]
\caption{FedRAV Algorithm}
\label{alg:frl}
% \small
\footnotesize
\KwIn{Regional structure  $\{ \mathcal{A}_k \}^{K}_{k=1}$, local epochs $\kappa_1$, interval of intra-region aggregation $\kappa_2$, learning rate $\eta$, $\eta_{\boldsymbol{v}}$, $\eta_{\boldsymbol{\varphi}}$.}

\SetKwFunction{CentralServerExecute}{CentralServerExecute}
\SetKwFunction{RegionUpdate}{RegionUpdate}
\SetKwFunction{AVUpdate}{AVUpdate}
\SetKwProg{Fn}{function}{:}{}
\SetKwProg{fn}{}{:}{}

\KwOut{Personalized vehicular models $\{\mathcal{W}_k\}^k_{k=1}$ and regional models ${\mathcal{W}}$.}

\Fn{\CentralServerExecute}
{   
    Initialize $\mathcal{W}^{(t_0)} = [\boldsymbol{\tilde{w}}^{(t_0)}_1, \cdots, \boldsymbol{\tilde{w}}^{(t_0)}_K]$
    
    \For{each round $t\in \{0,...,T-1\}$}
    {     
        \For{ each regional server $k=1,...,K$ \textbf{in parallel}}
        {
            $\boldsymbol{\tilde{w}}^{(t + 1)}_k \gets \boldsymbol{\tilde{w}}^{(t)}_k + (\mathcal{W}^{(t)} \setminus \boldsymbol{\tilde{w}}^{(t)}_k) \odot h_k(\boldsymbol{v}^{(t)}_k;\boldsymbol{\varphi}^{(t)}_k)$ 

            $\Delta \boldsymbol{w}_k = $ \RegionUpdate{$t$, $k$, $\boldsymbol{\tilde{w}}^{(t + 1)}_k, \mathcal{A}_k$}

            Update ${\mathcal{W}^{(t)}}$ according to $\Delta \boldsymbol{w}_k$

            $\boldsymbol{v}^{(t + 1)}_k \gets \boldsymbol{v}^{(t)}_k - \eta_{\boldsymbol{v}} \left( \nabla _{\boldsymbol{v}^{(t)}_k}\boldsymbol{\tilde{w}}^{(t+1)}_k \right) ^T \Delta \boldsymbol{w}_k$

            $\boldsymbol{\varphi}^{(t + 1)}_k \gets \boldsymbol{\varphi}^{(t)}_k - \eta_{\boldsymbol{\varphi}} \left( \nabla _{\boldsymbol{\varphi}^{(t)}_k}\boldsymbol{\tilde{w}}^{(t+1)}_k \right) ^T \Delta \boldsymbol{w}_k$
        }
    }

    \textbf{return} personalized regional models ${\mathcal{W}^{(T-1)}}$
}

\Fn{\RegionUpdate{$t$, $k$, $\boldsymbol{\tilde{w}}^{(t)}_k$, $\mathcal{A}_k$}}
{
     Initialize $\mathcal{W}^{(t)}_k=[\boldsymbol{\tilde{w}}^{(t)}_i]_{i \in \mathcal{A}_k}$ with $\boldsymbol{\tilde{w}}^{(t)}_k$
     
    \For{each AV $i \in \mathcal{A}_k$ \textbf{in parallel}}
    {
        $\boldsymbol{\tilde{w}}^{(t + 1)}_i \gets \boldsymbol{\tilde{w}}^{(t)}_i + (\mathcal{W}^{(t)}_k \setminus \boldsymbol{\tilde{w}}^{(t)}_i) \odot h_i(\boldsymbol{v}^{(t)}_i;\boldsymbol{\varphi}^{(t)}_i)$

        $\Delta \boldsymbol{w}_i = $ \AVUpdate{$t$, $k$, $i$, $\boldsymbol{\tilde{w}}^{(t + 1)}_i$}

        Update ${\mathcal{W}^{(t)}_k}$ according to $\Delta \boldsymbol{w}_i$

        $\boldsymbol{v}^{(t + 1)}_i \gets \boldsymbol{v}^{(t)}_i - \eta_{\boldsymbol{v}} \left( \nabla _{\boldsymbol{v}^{(t)}_i}\boldsymbol{\tilde{w}}^{(t+1)}_i \right) ^T \Delta \boldsymbol{w}_i$ \Comment{ Eq.\ref{eq:gradient_v_i}}

        $\boldsymbol{\varphi}^{(t + 1)}_i \gets \boldsymbol{\varphi}^{(t)}_i - \eta_{\boldsymbol{\varphi}} \left( \nabla _{\boldsymbol{\varphi}^{(t)}_i}\boldsymbol{\tilde{w}}^{(t+1)}_i \right) ^T \Delta \boldsymbol{w}_i$ \Comment{ Eq.\ref{eq:gradient_phi_i}}

    }

    \If{ $t\equiv 0 \pmod {\kappa_2}$}
   % $t \% \kappa_2 = 0$ }
    {
       Aggregation 
       $\boldsymbol{w}^{(t)}_k = \sum_{i \in \mathcal{A}_k} \frac{g\left( \lVert \tilde{w}^{(t)}_i-\bar{w}_{k} \rVert \right)}{\sum_{j\in \mathcal{A}_k}{g\left( \lVert \tilde{w}^{(t)}_j-\bar{w}_{k} \rVert \right)}} \tilde{w}^{(t)}_i$

        \textbf{return} $\Delta \boldsymbol{w}_k = \boldsymbol{w}^{(t)}_k - \boldsymbol{\tilde{w}}^{(t)}_k$ to central server
    }
}

\Fn{\AVUpdate{$t$, $k$, $i$, $\boldsymbol{\tilde{w}}^{(t)}_{i}$}}
{
    Initialize local model $\boldsymbol{w}^{(t)}_{i}=\boldsymbol{\tilde{w}}^{(t)}_{i}$

    \For{local epoch $e = 1,\cdots, \kappa_1$}
     {
        $\boldsymbol{w}^{(t)}_{i} \gets \boldsymbol{w}^{(t)}_{i} -\eta \nabla F_i\left(\boldsymbol{w}^{(t)}_{i}\right)$  \Comment{Local Update}

     }
                        
    \textbf{return}  $\Delta \boldsymbol{w}_i = \boldsymbol{w}^{(t)}_{i} - \boldsymbol{\tilde{w}}^{(t)}_{i}$ to regional server
}
%\vspace{-0.15in}
\end{algorithm}

\subsection{Personalization via Hypernetworks} \label{sec:personalization}

Inspired by \cite{
shamsian2021personalized}
% huang2021personalized,
% , ma2022layer}
, a personalized model can be viewed as the linear combination of model parameters from other AVs. Due to the discrepancy of model parameter, we aim to  
estimate the proper linear combination coefficients via hypernetworks.
%how can we estimate the proper linear combination coefficients? 
%We handle this problem via hypernetworks. 
Suppose that hypernetwork $h(\cdot;\cdot)$ can be parameterized by the embedding vector $\boldsymbol{v}$ and model parameter $\boldsymbol{\varphi}$, i.e., $h_i(\boldsymbol{v}_i;\boldsymbol{\varphi}_i)$ denotes hypernetwork designated by AV $i$ and $h_{k}(\boldsymbol{v}_{k}; \boldsymbol{\varphi}_{k})$ denotes hypernetwork related to region $k$. Regional server $k$ holds a set of model parameters of AVs $\mathcal{W}_k=[\boldsymbol{\tilde{w}}_i]_{i \in \mathcal{A}_k}$ and central server has a set of model parameters of regions $\mathcal{W}=[\boldsymbol{\tilde{w}}_k]^K_{k=1}$. Accordingly, personalized vehicular model of AV $i$ is defined as follows,
\begin{equation}
\small
    \boldsymbol{w}_i = \tilde{\boldsymbol{w}}_i+ (\mathcal{W}_{k} \setminus \tilde{\boldsymbol{w}}_i)\odot h_i(\boldsymbol{v}_i;\boldsymbol{\varphi}_i).
\end{equation}
% [zz]%Fig.3里面setminus符号右边的部分记得像上面一样，加一个中括号，不然可能有歧义
The first term $\boldsymbol{\tilde{w}}_i$ is the model parameter from AV $i$ that helps maintain the personalized part of the model. The second term represents empirical knowledge transferred from other AVs' models. The notation $\odot$ calculates the inner product between the models and mask vectors. 
For simplicity, let $\boldsymbol{\alpha}_i = h_i(\boldsymbol{v}_i; \boldsymbol{\varphi}_i)$ which is called personalized mask vectors satisfying $\sum_{j \in \{ \mathcal{A}_{k} \setminus i \}} \boldsymbol{\alpha}_{ij}=1$, i.e. the summation of all entries of $\boldsymbol{\alpha}_i$ equals $1$. 
Each element of $\boldsymbol{\alpha}_i$ is in $[0, 1]$.  Thus, $\boldsymbol{w}_i$ can be writed as,
\begin{equation}
\small
    \boldsymbol{w}_i = \tilde{\boldsymbol{w}}_i+ \sum_{j \in \{ \mathcal{A}_{k} \setminus i \}} \tilde{\boldsymbol{w}}_j \boldsymbol{\alpha}_{ij}.
\end{equation}
Specifically, the larger value of $\boldsymbol{\alpha}_{ij}$ represents the greater contribution to the personalized model. 
%On the contrary, FRL 
Vice versa, FedRAV
may discard the unprofitable models by setting the corresponding values $\boldsymbol{\alpha}_{ij}$ to zero. 
The adaptive mask vectors $\boldsymbol{\alpha}_i$ allow knowledge transfer across AVs or regions while alleviating the disturbance of unprofitable models.

Furthermore, we can readjust the optimization objective Eq. \ref{eqn:p2} according to the above analysis as follows,
\begin{equation}
\small
    \{(\boldsymbol{v}_{i}^{*},\boldsymbol{\varphi }_{i}^{*}),\cdots \}=
    \argmin_{(\boldsymbol{v}_i,\boldsymbol{\varphi }_i),\cdots}
\frac{1}{\left| \mathcal{A}_k \right|}\sum_{i\in \mathcal{A}_k}{F_i}\left(\mathcal{D}_i; \boldsymbol{w_i}\right),
    \label{eqn:new_p2}
\end{equation}
where $\boldsymbol{w}_i= \boldsymbol{\tilde{w}}_i + (\mathcal{W}_k \setminus \boldsymbol{\tilde{w}}_i)   \odot h_i\left( \boldsymbol{v}_i;\boldsymbol{\varphi }_i \right)$. Similarly, the optimization objective Eq. \ref{eqn:p3} can be rewritten as follows,
\begin{equation}
\small
    \{(\boldsymbol{v}_{k}^{*},\boldsymbol{\varphi }_{k}^{*}),\cdots \} =
    \argmin_{(\boldsymbol{v}_k, \boldsymbol{\varphi }_k),\cdots}\frac{1}{K}\sum_{k=1}^K{f_k}\left(\mathcal{D}_k; \boldsymbol{w}_k \right), 
    \label{eqn:new_p3}
\end{equation}
where $ \boldsymbol{w}_k =  \boldsymbol{\tilde{w}}_k +  (\mathcal{W} \setminus \boldsymbol{\tilde{w}}_k) \odot h_k\left( \boldsymbol{v}_k;\boldsymbol{\varphi}_k \right)$. Therefore, we transform the optimization of $\mathcal{W}_k$ and $\mathcal{W}$ to optimization of  $\{(\boldsymbol{v}_{i},\boldsymbol{\varphi }_{i}),\cdots\}$ and $\{(\boldsymbol{v}_{k},\boldsymbol{\varphi }_{k}),\cdots \}$, respectively. By using the chain rule, the gradient of Eq. \ref{eqn:new_p2} is derived with respect to the optimization variable $\boldsymbol{v}_i$ as follows,
\begin{equation}
\small
% \begin{aligned}
    \nabla_{\boldsymbol{v}_i}F_i
    =\left( \nabla _{\boldsymbol{v}_i}\boldsymbol{w}_i \right) ^T\nabla _{\boldsymbol{w}_i}F_i 
    =\left[(\mathcal{W}_{k} \setminus \boldsymbol{\tilde{w}}_i) \odot \nabla _{\boldsymbol{v}_i}h_i \right] ^T\nabla _{\boldsymbol{w}_i}F_i.
% \end{aligned}
\label{eq:gradient_v_i}
\end{equation}
For $\boldsymbol{\varphi}_i$, we also have
\begin{equation}
\small
% \begin{aligned}
    \nabla _{\boldsymbol{\varphi}_i}F_i
    =\left( \nabla _{\boldsymbol{\varphi}_i}\boldsymbol{w}_i \right) ^T\nabla _{\boldsymbol{w}_i}F_i
    =\left[ (\mathcal{W}_{k} \setminus \boldsymbol{\tilde{w}}_i) \odot \nabla _{\boldsymbol{\varphi}_i}h_i\right] ^T\nabla _{\boldsymbol{w}_i}F_i.
% \end{aligned}
\label{eq:gradient_phi_i}
\end{equation}
Similarly, the gradient of Eq.\ref{eqn:new_p3} is obtained for $\boldsymbol{v}_k$ and $\boldsymbol{\varphi}_k$ as $\nabla_{\boldsymbol{v}_{k}}f_{k}=\left( \nabla _{\boldsymbol{v}_{k}}\boldsymbol{w}_{k} \right) ^T\nabla _{\boldsymbol{w}_{k}}f_{k}$ and $\nabla_{\boldsymbol{\varphi}_{k}}f_{k} =\left( \nabla _{\boldsymbol{\varphi}_{k}}\boldsymbol{w}_{k} \right) ^T\nabla _{\boldsymbol{w}_{k}}f_{k}$, respectively. Afterwards, FedRAV can update the embedding vector and model parameter of the hypernetwork via gradient descent. Note that, we simplify the computation of gradient $\nabla _{\boldsymbol{w}_i}F_i$ and $\nabla _{\boldsymbol{w}_k}f_k$ and replace them with pseudo-gradient $\Delta \boldsymbol{w}_i$ and $\Delta \boldsymbol{w}_k$ following \cite{shamsian2021personalized}
% , ma2022layer}
as shown in Algorithm \ref{alg:frl}. Our experiments (Section \ref{sec:experiments}) prove that the pseudo-gradient is highly efficient.
In practice, FedRAV can execute the backpropagation of hypernetwork to obtain the gradients $\nabla _{\boldsymbol{v}_{i}}\boldsymbol{w}_{i}$, $\nabla _{\boldsymbol{\varphi}_{i}}\boldsymbol{w}_{i}$, $\nabla _{\boldsymbol{v}_{k}}\boldsymbol{w}_{k}$ and $\nabla _{\boldsymbol{\varphi}_{k}}\boldsymbol{w}_{k}$.

% \vspace{-0.1in}

\subsection{Intra-region Aggregation Policy} \label{sec:aggregation}

In order to optimize the personalized regional model, we propose a heuristic intra-region aggregation policy with a penalty function. $\mathcal{W}_k=[\boldsymbol{\tilde{w}}_i]_{i \in \mathcal{A}_k}$ denotes the set of models from AVs belonging to region $k$. $\bar{\boldsymbol{w}}_k = \frac{1}{\vert \mathcal{A}_k \vert}\sum_{i \in \mathcal{A}_k} \boldsymbol{\tilde{w}}_i$ denotes the average model of region $k$. Instead of average model $\bar{\boldsymbol{w}}_k$,  the model that reflects the preferences of the majority of AVs in the region $k$ is also required, which is described as $\textit{regional model}$. Specifically, the average model parameter $\bar{ \boldsymbol{w}}_k$ is selected as the reference model parameter in the regional model parameter space. $\Vert \boldsymbol{\tilde{w}}_i - \boldsymbol{\bar{w}}_k \Vert$ represents the norm distance between $\boldsymbol{\tilde{w}}_i$ and average model $\boldsymbol{\bar{w}}_k$. Intuitively, the regional model 
%should be able to reflects the preferences of 
prefers models closer to the average model.

% \begin{figure*}
%      \centering
%      \subfigure{
%          \includegraphics[width=1.9in]{gtsrb_p2_.pdf}
%         \includegraphics[width=1.9in]{miotcd_p2_.pdf}
%         \includegraphics[width=1.9in]{vehicle-10_p2_.pdf}
%      }
%      {\hspace{1.1in} (a) \hspace{1.73in} (b) \hspace{1.73in} (c) \hspace{-0.22in}}
%      \vspace{-0.1in}
%      \caption{Comparison of different FL approaches in $\rho=20\%$ Non-IID setting (a) GTSRB; (b) MIO-TCD; (c) Vehicle-10.}
%      \label{fig:compar_p2}
%      \vspace{-0.15in}
% \end{figure*}

% \begin{figure*}
%      \centering
%     \subfigure {
%         \includegraphics[width=1.9in]{gtsrb_p3_.pdf}
%         \includegraphics[width=1.9in]{miotcd_p3_.pdf}
%         \includegraphics[width=1.9in]{vehicle-10_p3_.pdf}
%     }
%     {\hspace{1.1in} (a) \hspace{1.73in} (b) \hspace{1.73in} (c) \hspace{-0.22in}}
%      \vspace{-0.1in}
%      \caption{Comparison of different FL approaches in $\rho=30\%$ Non-IID setting (a) GTSRB; (b) MIO-TCD; (c) Vehicle-10.}
%      \label{fig:compar_p3}
%      \vspace{-0.15in}
% \end{figure*}

According to the above analysis, the intra-region aggregation policy is formally proposed as follows,
\begin{equation}
\small
\boldsymbol{w}_k = \sum_{i \in \mathcal{A}_k}  \frac{g\left( \lVert \tilde{\boldsymbol{w}}_i-\bar{\boldsymbol{w}}_{k} \rVert \right)}{\sum_{j\in \mathcal{A}_k}{g\left( \lVert \tilde{\boldsymbol{w}}_j-\bar{\boldsymbol{w}}_{k} \rVert \right)}} \tilde{\boldsymbol{w}}_i,
\end{equation}
where $g(\cdot)= e^{-(\cdot)}$ is a penalty function. Function $g(\cdot)$ is a monotonically decreasing function defined on $[0, +\infty]$ with the value ranges $[0,1]$. For simplicity, let $\beta_i = \frac{g\left( \lVert \tilde{\boldsymbol{w}}_i-\bar{\boldsymbol{w}}_{k} \rVert \right)}{\sum_{j\in \mathcal{A}_k}{g\left( \lVert \tilde{\boldsymbol{w}}_j-\bar{\boldsymbol{w}}_{k} \rVert \right)}}$ which represents 
%the size of 
the contribution of $\tilde{\boldsymbol{w}}_i$ to the regional model. 
%Precisely, if the personalized vehicular model $\boldsymbol{\tilde{w}}_i$ is further away from the average model $\bar{\boldsymbol{w}}_k$, the penalty function $g$ will impose a greater penalty, making the aggregate weight $\beta_i$ smaller. 
Specifically,  $g(\cdot)$ will impose a greater penalty if the personalized vehicular model $\boldsymbol{\tilde{w}}_i$ is further from the average model $\bar{\boldsymbol{w}}_k$. Therefore, lower aggregation weight is assigned to model $\boldsymbol{\tilde{w}}_i$.
Then, the region $k$ can compute the pseudo-gradient by $\Delta \boldsymbol{w}_k = \boldsymbol{w}_k - \boldsymbol{\tilde{w}}_k$ and upload it to the central server for updating hypernetwork, as shown in Algorithm \ref{alg:frl}.

% In summary, our unifying framework 

% \begin{figure}
%      \centering
%     \includegraphics[width=2.5in]{dataset.pdf}
%     {\hspace{0.7in} (a) GTSRB   \hspace{0.15in}  (b) MIO-TCD  \hspace{0.12in}  (c) Vehicle-10 \hspace{-0.15in}}
%      \vspace{-0.1in}
%      \caption{The samples of three real world  datasets.}
%      \label{fig:dataset}
%      \vspace{-0.2in}
% \end{figure}

\begin{figure}
     \centering
     \subfigure{
        \centering
        \includegraphics[width=1.65in]{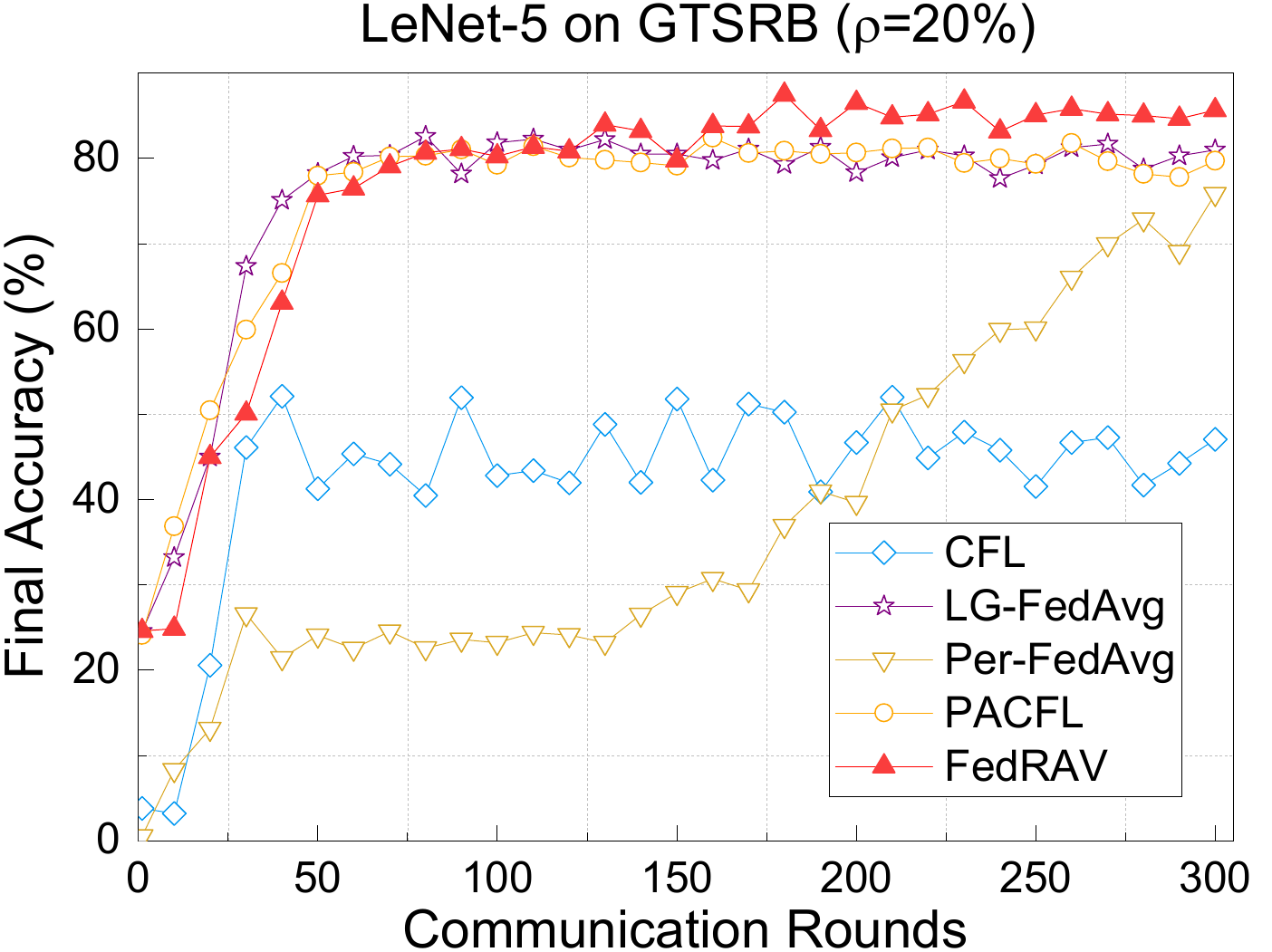}
        \includegraphics[width=1.65in]{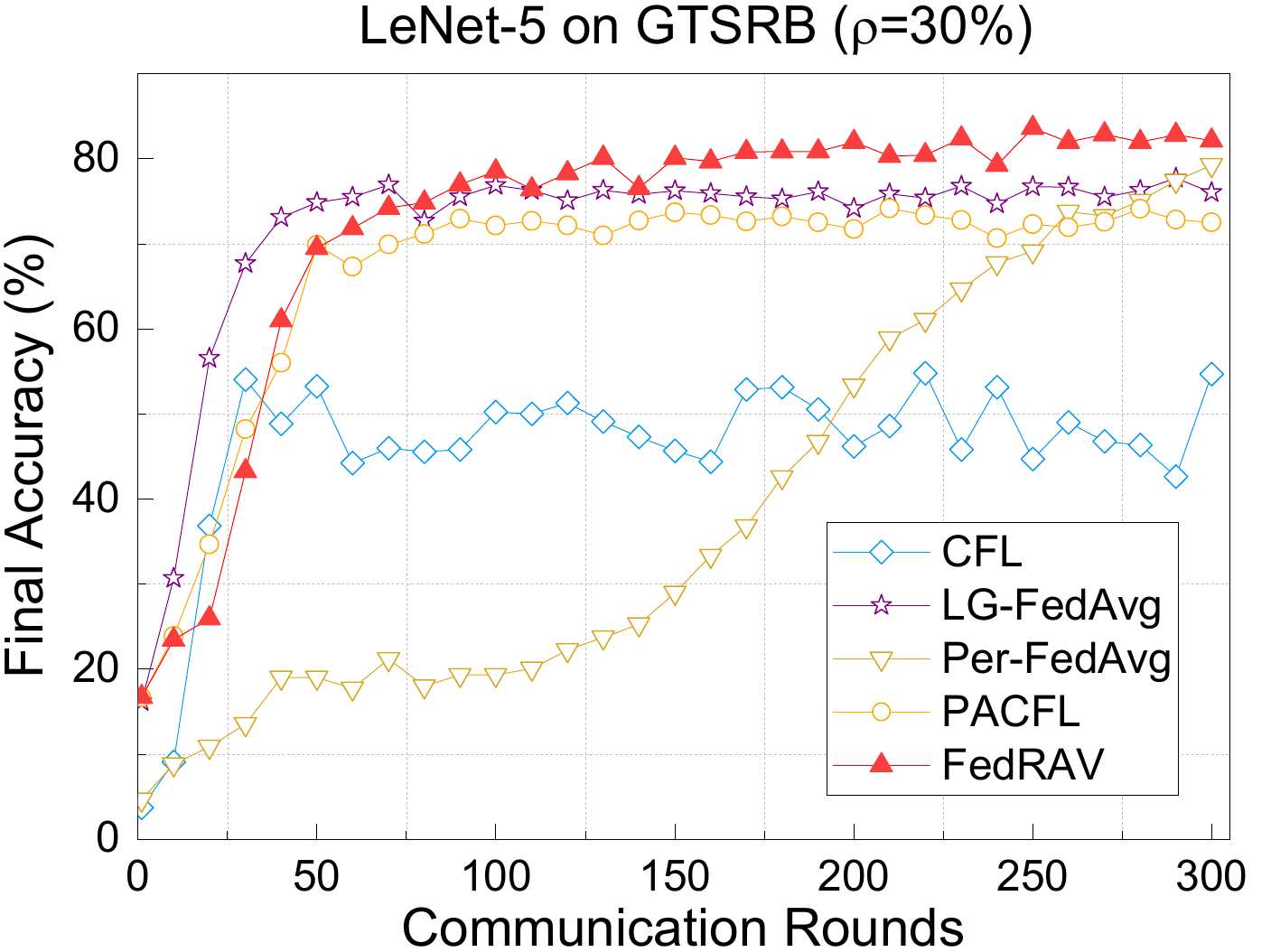}
     }
     {\small  \hspace{0.38in}  (a) GTSRB ($\rho=20\%$) \hspace{0.43in}  (b) GTSRB ($\rho=30\%$) \hspace{-0.82in} }
     \vspace{0.05in}
     \subfigure{
        \centering
        \includegraphics[width=1.65in]{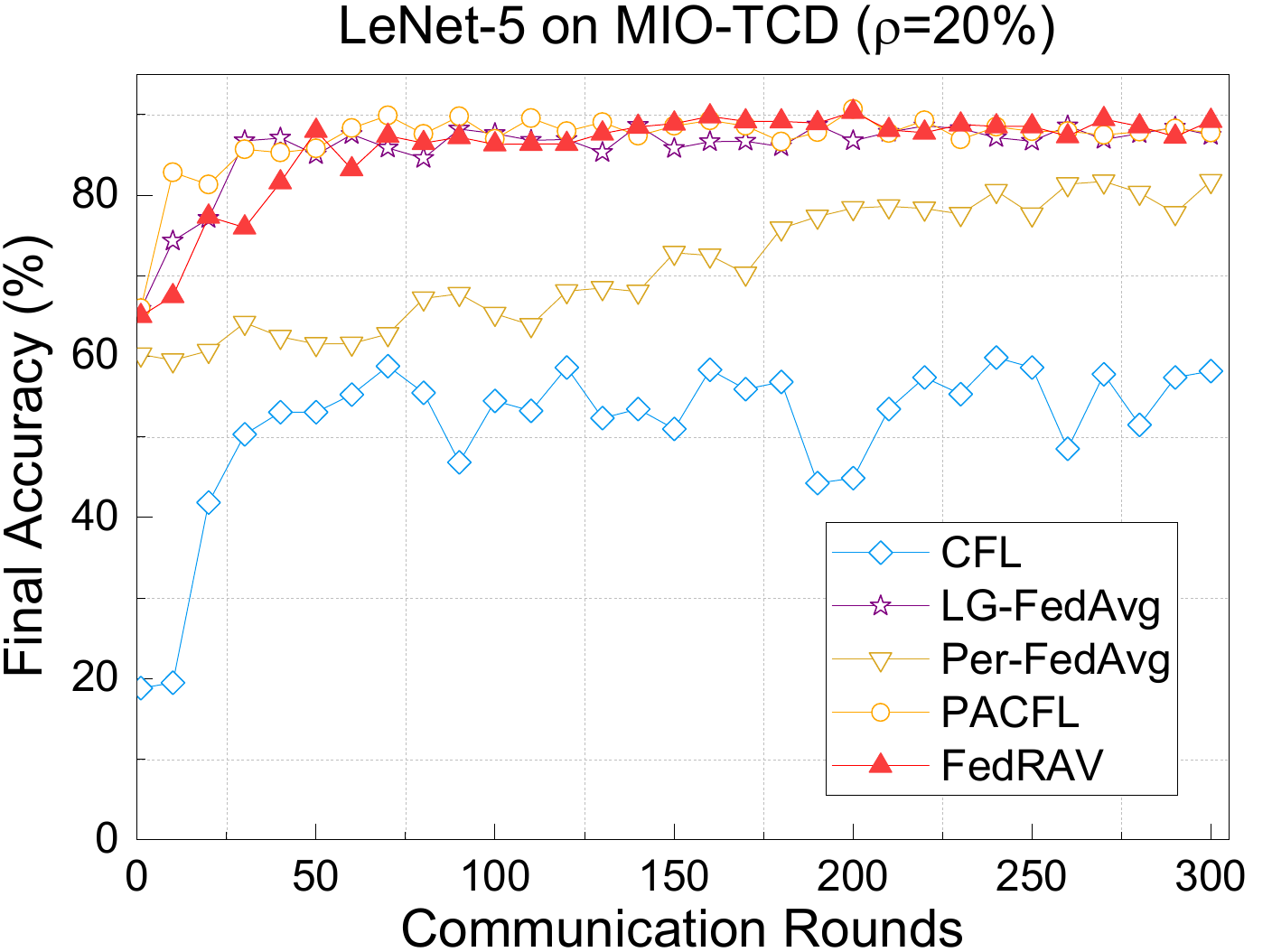}
        \includegraphics[width=1.65in]{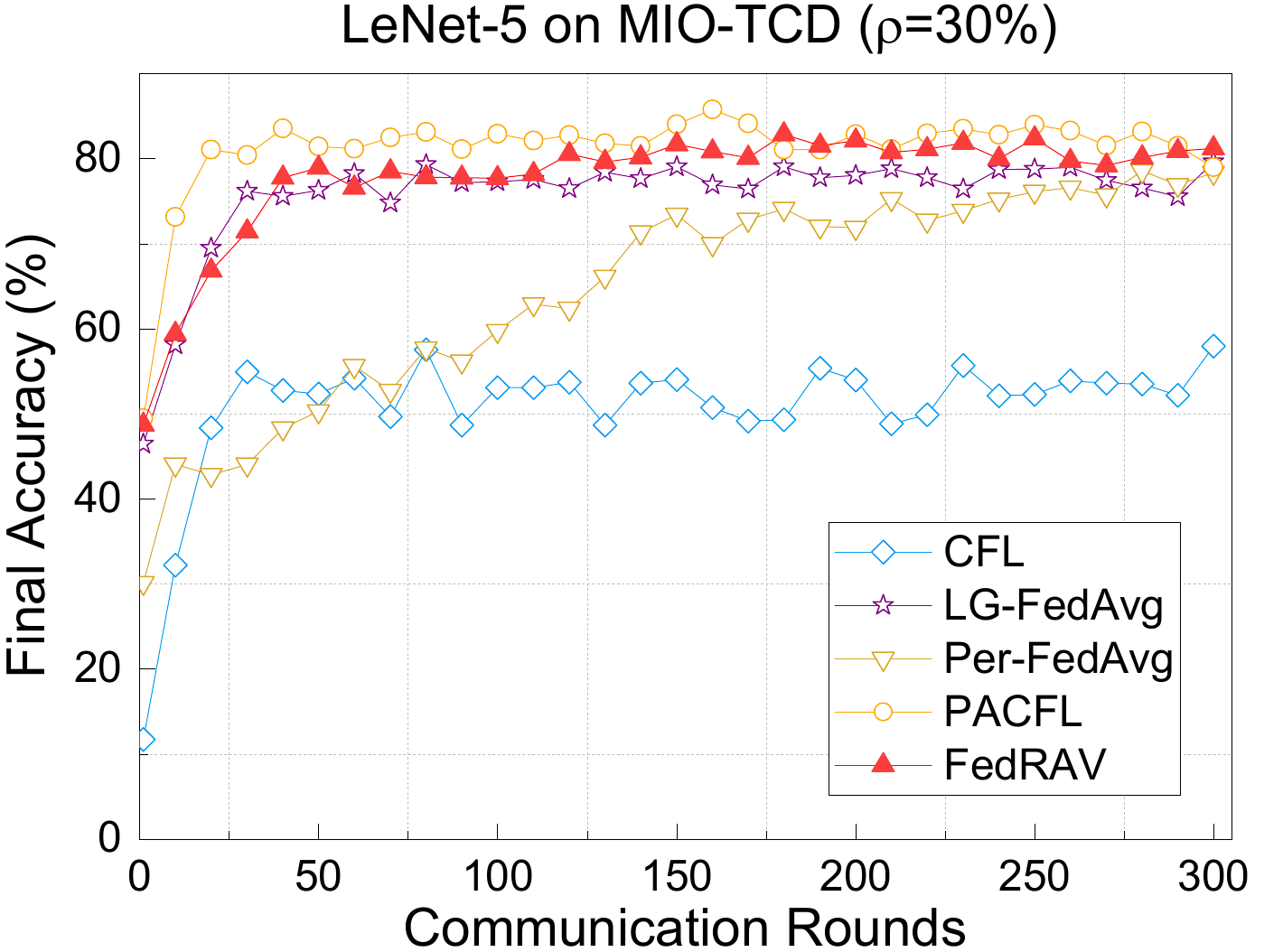}
     }
     {\small  \hspace{0.48in}  (c) MIO-TCD ($\rho=20\%$) \hspace{0.33in}  (d) MIO-TCD ($\rho=30\%$) \hspace{-0.82in} }
     \vspace{0.05in}
     \subfigure{
        \centering
        \includegraphics[width=1.65in]{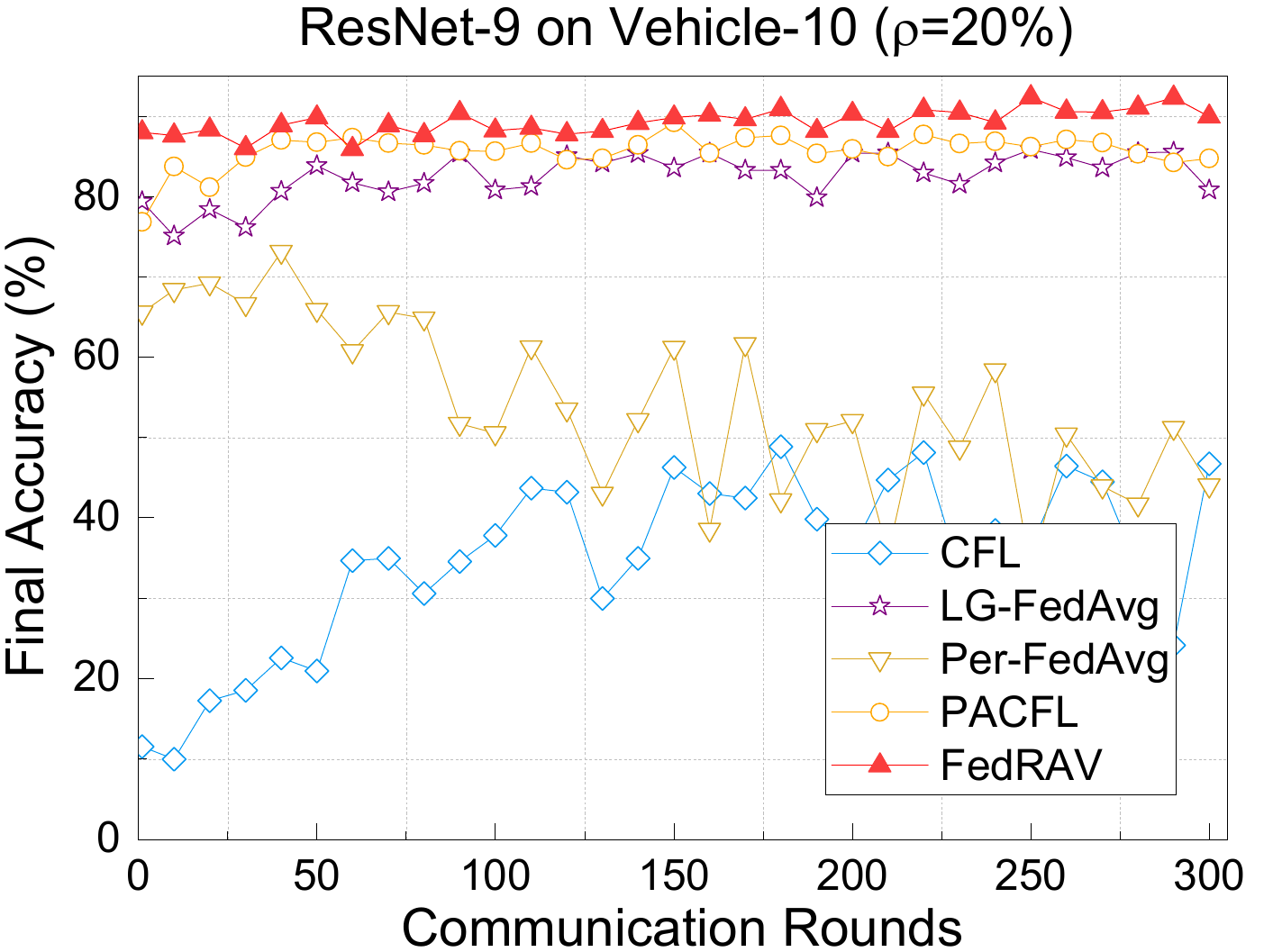}
        \includegraphics[width=1.65in]{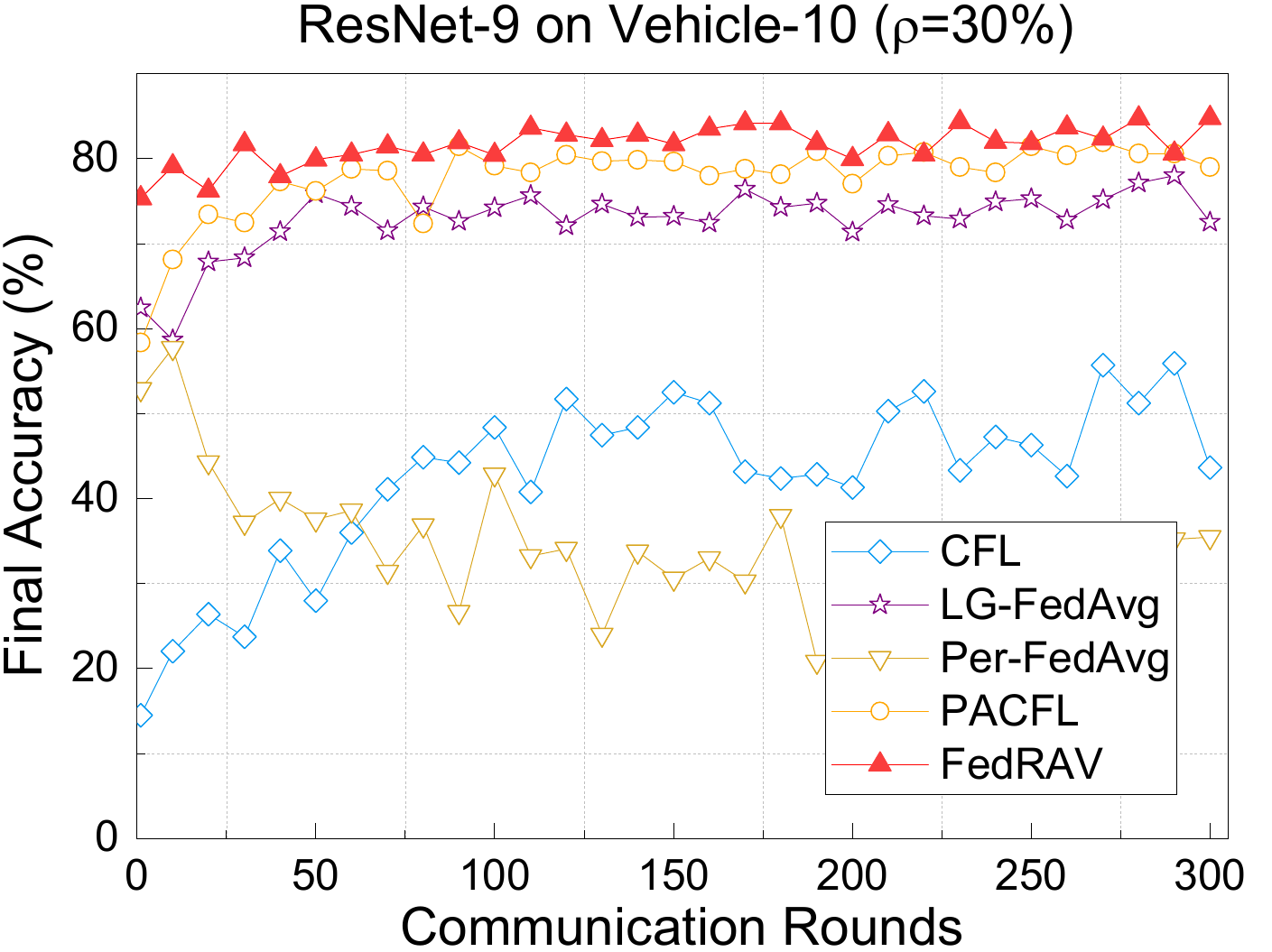}
     }
      {\small  \hspace{0.48in}  (e) Vehicle-10 ($\rho=20\%$) \hspace{0.33in}  (f) Vehicle-10 ($\rho=30\%$) \hspace{-0.12in} }
     \caption{Comparison of different FL approaches.
     }
     \label{fig:compar_p2_p3}
     \vspace{-0.20in}
\end{figure}

\section{Experiments}  \label{sec:experiments}
\subsection{Experimental Setup} \label{sec:setup}

% [zz]%一定在显著位置注明，代码会开源，因为匿名的关系，可以先不放链接。如果要放链接，那代码链接不能违反匿名要求。
\textbf{Datasets and Models.}
We consider 
%image classification task 
traffic object classification task
and evaluate our framework in three real-world datasets: GTSRB\cite{Houben-IJCNN-2013}, MIO-TCD\cite{luo2018mio} and Vehicle-10, which are widely used to evaluate autonomous driving algorithms. 
GTSRB, the German traffic sign recognition benchmark, consists of 39270 images of 43 different traffic signs.
% , as shown in Fig. \ref{fig:dataset}(a).
The classification section of MIO-TCD contains a total of 52801 images from 11 categories of traffic participants.
% , as shown in Fig. \ref{fig:dataset}(b).
In this work, we also collected 36006 vehicle images from the Internet and divided them into ten categories, which we called Vehicle-10\footnote{Vehicle-10 is open-accessed at: https://github.com/yjzhai-cs/Vehicle-10.}.
% , 
% \footnote{Vehicle-10 dataset is open sourced at: anonymous link. }
% as shown in Fig. \ref{fig:dataset}(c).
For all experiments, we consider LetNet-5\cite{lecun1989backpropagation} for the GTSRB and MIO-TCD datasets, and ResNet-9\cite{he2016deep} for the Vehicle-10.
% dataset.
% Specifically, all of these images are resized to $32$pix $\times$ $32$pix.

% Unfortunately, these datasets do not provide meta information of vehicle GPS. 

% \textbf{Compared Baselines.} First, we evaluate the effectiveness of our FedRAV framework. Then, we compare and evaluate the following single-model-based algorithms. 1) \textbf{FedAvg} \cite{mcmahan2017communication} is the first proposed federated learning algorithm. 2) \textbf{FedProx} \cite{li2020federated} introduces regularization terms into the local objective function to control the deviation of the local model from the global model.
% % , which effectively alleviates the influence of Non-IID data. 
% 3) \textbf{FedNova} \cite{wang2020tackling} uses a standard normalization approach to eliminate target inconsistencies while maintaining fast convergence. In addition, we also evaluate several multi-model approaches.
% 4) \textbf{CFL} \cite{sattler2020clustered} iteratively divides all clients according to the cosine similarity based on gradient update so that the dissimilarity of inter-clusters distributions is as large as possible, which mitigates the impact of heterogeneous model aggregation. 5) \textbf{LG-FedAvg} \cite{liang2020think} combines local representation learning with global model learning, which reduces variance in the data. 6) \textbf{Per-FedAvg} \cite{fallah2020personalized} implements personalization by meta-learning approach. 7) \textbf{PACFL} \cite{vahidian2023efficient} identifies distribution similarity to form clusters by analyzing the principal angles between client data subspaces.

\textbf{Compared Baselines.} First, we evaluate the effectiveness of our FedRAV framework. Then, we compare and evaluate the following single-model algorithms: FedAvg \cite{mcmahan2017communication}, FedProx \cite{li2020federated}, and FedNova \cite{wang2020tackling}. In addition, we also evaluate several multi-model approaches: CFL \cite{sattler2020clustered}, LG-FedAvg \cite{liang2020think}, Per-FedAvg \cite{fallah2020personalized}, and PACFL \cite{vahidian2023efficient}.

\textbf{Implementation Details.} In all experiments, we suppose that 100 AVs participate in learning tasks. Unfortunately, the above three datasets do not provide metadata of vehicle GPS. We randomly generate location coordinates of 100 AVs for each dataset, ensuring that vehicles in proximity have comparatively similar data distributions. 
Note that this synthetic coordinate information of AV is only a supplement. In order to simulate the Non-IID label skewed setting, we randomly assign $\rho$ of the total labels of the dataset to each AV, i.e., $\rho=20\%$ or $\rho=30\%$. For all experiments, The number of training rounds is set to $T=300$, and $20\%$ of the AVs are sampled randomly to participate in the training process at each round. AV trains the local model by stochastic gradient descent with iteration $\kappa_1=10$ and batch-size $B=20$. 
% Specifically, the learning rate used for updating parameters of the hypernetwork is $\eta_{\boldsymbol{\varphi}} = 0.001$, and the learning rate used for updating embedded vectors is $\eta_{\boldsymbol{v}}=0.005$. Without loss of generality, a multilayer perceptron model with two hidden layers is chosen as the hypernetwork. All code is implemented in Pytorch and runs on $8 \times$ NVIDIA GeForce RTX $3090$ GPUs. 

\begin{table*}[t]
% \small
\footnotesize
\centering
\caption{Quantitative comparison results across different datasets in ($\rho=20\%$ and $\rho=30\%$) Non-IID settings. }
\vspace{-0.05in}
\begin{tabular}{c|c|c|c|c|c|c|c}
\toprule
\multicolumn{2}{c|}{Dataset} & \multicolumn{2}{c|}{GTSRB} &\multicolumn{2}{c|}{MIO-TCD} &\multicolumn{2}{c}{Vehicle-10}\\
% \hline
\midrule
\multicolumn{2}{c|}{ Non-IID level $\rho$}  & $20\%$ & $30\%$ & $20\%$ & $30\%$ & $20\%$ & $30\%$ \\
\midrule

%single model-----------------
\multirow{3}{*}{Single-Model} & FedAvg\cite{mcmahan2017communication} & $84.98 \pm 0.31$ & $84.40 \pm 1.87$ &
$63.93\pm1.88$ & $66.19\pm0.94$ & $62.18\pm4.03$ &	$69.33\pm6.04$ \\

\multirow{3}{*}{} & FedProx\cite{li2020federated} & $83.97\pm0.95$ & $\textbf{85.18}\pm\textbf{0.16}$ & $64.25\pm0.88$
 & $67.90\pm1.04$  & $56.03\pm6.73$  & $69.67\pm6.83$	 \\

\multirow{3}{*}{} & FedNova\cite{wang2020tackling} & $83.18\pm1.05$  & $85.13\pm0.98$ & $63.70\pm3.66$
 & $67.57\pm1.29$ & $48.69\pm8.47$  & $68.65\pm5.08$	 \\
\midrule

%multi-model-----------------
\multirow{5}{*}{Multi-Model} & CFL\cite{sattler2020clustered} & $46.40\pm1.34$  & $53.34\pm1.24$  & $57.16\pm0.95$
 & $52.83\pm4.50$ & $46.27\pm1.01$ & $44.92\pm1.51$ \\

\multirow{5}{*}{} & LG-FedAvg\cite{liang2020think} & $80.55\pm0.69$  & $74.67\pm1.20$  & $86.42\pm0.92$
 & $77.96\pm1.63$ & $81.69\pm1.67$ & $74.66\pm1.84$\\

\multirow{5}{*}{} & Per-FedAvg\cite{fallah2020personalized} & $73.04\pm4.82$  & $78.23\pm0.94$  & $79.71\pm1.81$
 & $77.89\pm0.90$ & $43.58\pm13.2$ & $31.78\pm4.61$ \\

\multirow{5}{*}{} & PACFL\cite{vahidian2023efficient} & $79.31\pm0.49$  & $71.44\pm0.94$  & $88.44\pm1.03$
 & $81.08\pm1.90$ & $86.43\pm1.80$ & $78.11\pm0.85$ \\

\multirow{5}{*}{} & \textbf{FedRAV} & $\textbf{86.55}\pm\textbf{0.83}$  & $83.37\pm1.21$  & $\textbf{88.72}\pm\textbf{0.99}$
 & $\textbf{81.65}\pm\textbf{0.49}$ & $\textbf{89.77}\pm\textbf{0.64}$ & $\textbf{84.02}\pm\textbf{0.67}$ \\

\bottomrule
\end{tabular}

\label{table:compar}
\vspace{-0.05in}
\end{table*}

\textbf{Evaluation Metrics.} For single-model approaches, we evaluate the global model on the test set and report classification accuracy as experiment results. For multi-model approaches, each client holds a local test set in our experimental setting. We evaluate the personalized vehicular model on the local test set and use the average of final local test accuracy (a.k.a. final accuracy) to measure the performance of FL approaches.

\subsection{Performance Evaluation}
% We first evaluate the performance of FedRAV on the GTSRB, MIO-TCD, and Vehicle-10 datasets against the above baselines.
% single-model baselines (FedAvg, FedProx, and FedNova) and multi-model baselines (CFL, LG-FedAvg, Per-FedAvg, and PACFL). 
% The results of final accuracy in each communication round are shown in Fig. \ref{fig:compar_p2} and Fig \ref{fig:compar_p3}. We also report the mean and standard deviation for the final accuracy over $3$ runs as shown in Table \ref{table:compar}. 
% The experiment results show that FedRAV is superior to other baselines.
% and improves final accuracy by $+39.24\%$ on GTSRB, $+32.05\%$ on MIO-TCD, $+43.71\%$ on Vehicle-10 for CFL in $\rho=20\%$ setting.
% [zz]%为什么表三里只有FedRAV方法是+-0？

% \textbf{Effectiveness of FedRAV.} 
As shown in Fig. \ref{fig:compar_p2_p3}, we observe that the FedRAV achieves the superior performance and outperforms the strong competitors on the most datasets with $K=5$, $\kappa_1=10$, $\kappa_2=10$, and $\gamma=0.5$. Besides, the CFL achieves the worst convergence on GTSRB and MIO-TCD datasets compared to other baselines, which owes to its failed clustering strategy. For example, CFL may partition a single AV into a separate cluster, leading to inadequate training. However, FedRAV can resolve this situation by gathering nearby AVs into a trainable region. In addition, Per-FedAvg converges with lower accuracy than other frameworks on the Vehicle-10 dataset. This is interpreted as highly Non-IID data that makes the approach suffer the suboptimal solution, i.e., optimization direction gradually deviates from the optimal solution. The accuracy curve illustrates that FedRAV can converge with similar accuracy as LG-FedAvg and PACFL before $100$ communication rounds. Henceforth, FedRAV achieves higher accuracy, whereas LG-FedAvg and PACFL converge at a lower accuracy. Therefore, FedRAV achieves comparable convergence and superior accuracy in Non-IID setting and does not induce more communication rounds in federated learning systems.

Table \ref{table:compar} presents the final accuracy over all baselines on three datasets. We find that most multi-model methods (LG-FedAvg, PACFL, and ours FedRAV) outperform the single-model methods on highly heterogeneous MIO-TCD and Vehicle-10 datasets, which further confirms the 
%mentioned earlier hypothesis
claim that the single model failed to fit heterogeneous client's model parameter spaces. For the GTSRB dataset, single-model approaches converge the comparable accuracy with multi-model approaches. This is because the GTSRB dataset 
%with $43$ categories in $\rho=20\%$ and $\rho=30\%$ setting 
has weaker heterogeneity than the MIO-TCD and Vehicle-10. In particular, FedProx is slightly ahead of FedAvg on most datasets.
% The observation implies that the trick of adding a regularization term to the local objective function does not effectively mitigate data heterogeneity.
FedRAV achieves superior performance and outperforms most of the state-of-the-art methods. Concretely, FedRAV improves final accuracy by $+6\%$ on GTSRB, $+2.3\%$ on MIO-TCD and $+8.08\%$ on Vehicle-10 for LG-FedAvg in $\rho=20\%$ setting, and 
enhances final accuracy by $+8.7\%$ on GTSRB, $+3.69\%$ on MIO-TCD and $+9.36\%$ on Vehicle-10 for LG-FedAvg in $\rho=30\%$ setting. 

\section{Conclusion}  \label{sec:conclusion}
In this work, we propose a two-stage framework for learning personalized driving models from the scattered data collected from vehicles in various regions. It contains an efficient partitioning mechanism with one-shot communication for dividing large areas into sub-regions, and a FedRAV framework to employ the hypernetworks to learn personalized driving models against the heterogeneity, facilitating knowledge transfer among clients and regions.
% In particular, we theoretically prove that the partitioning mechanism can get a good initialization with $O(logK)$-approximation to the global optimum in expectation with polynomial complexity.
%and also analyze the complexity of the proposed mechanism. 
Finally, the extensive evaluations are conducted to verify the method's effectiveness and superior performance compared to state-of-the-art methods.

\begin{thebibliography}{10}
\providecommand{\url}[1]{#1}
\csname url@samestyle\endcsname
\providecommand{\newblock}{\relax}
\providecommand{\bibinfo}[2]{#2}
\providecommand{\BIBentrySTDinterwordspacing}{\spaceskip=0pt\relax}
\providecommand{\BIBentryALTinterwordstretchfactor}{4}
\providecommand{\BIBentryALTinterwordspacing}{\spaceskip=\fontdimen2\font plus
\BIBentryALTinterwordstretchfactor\fontdimen3\font minus \fontdimen4\font\relax}
\providecommand{\BIBforeignlanguage}[2]{{%
\expandafter\ifx\csname l@#1\endcsname\relax
\typeout{** WARNING: IEEEtran.bst: No hyphenation pattern has been}%
\typeout{** loaded for the language `#1'. Using the pattern for}%
\typeout{** the default language instead.}%
\else
\language=\csname l@#1\endcsname
\fi
#2}}
\providecommand{\BIBdecl}{\relax}
\BIBdecl

\bibitem{mcmahan2017communication}
B.~McMahan, E.~Moore, D.~Ramage, S.~Hampson, and B.~A. y~Arcas, ``Communication-efficient learning of deep networks from decentralized data,'' in \emph{Artificial intelligence and statistics}.\hskip 1em plus 0.5em minus 0.4em\relax PMLR, 2017, pp. 1273--1282.

\bibitem{fallah2020personalized}
A.~Fallah, A.~Mokhtari, and A.~Ozdaglar, ``Personalized federated learning: A meta-learning approach,'' \emph{arXiv preprint arXiv:2002.07948}, 2020.

\bibitem{sattler2020clustered}
F.~Sattler, K.-R. M{\"u}ller, and W.~Samek, ``Clustered federated learning: Model-agnostic distributed multitask optimization under privacy constraints,'' \emph{IEEE transactions on neural networks and learning systems}, vol.~32, no.~8, pp. 3710--3722, 2020.

\bibitem{vahidian2023efficient}
S.~Vahidian, M.~Morafah, W.~Wang, V.~Kungurtsev, C.~Chen, M.~Shah, and B.~Lin, ``Efficient distribution similarity identification in clustered federated learning via principal angles between client data subspaces,'' in \emph{Proceedings of the AAAI Conference on Artificial Intelligence}, vol.~37, no.~8, 2023, pp. 10\,043--10\,052.

\bibitem{zhao2018federated}
Y.~Zhao, M.~Li, L.~Lai, N.~Suda, D.~Civin, and V.~Chandra, ``Federated learning with non-iid data,'' \emph{arXiv preprint arXiv:1806.00582}, 2018.

\bibitem{li2020federated}
T.~Li, A.~K. Sahu, M.~Zaheer, M.~Sanjabi, A.~Talwalkar, and V.~Smith, ``Federated optimization in heterogeneous networks,'' \emph{Proceedings of Machine learning and systems}, vol.~2, pp. 429--450, 2020.

\bibitem{wang2020tackling}
J.~Wang, Q.~Liu, H.~Liang, G.~Joshi, and H.~V. Poor, ``Tackling the objective inconsistency problem in heterogeneous federated optimization,'' \emph{Advances in neural information processing systems}, vol.~33, pp. 7611--7623, 2020.

\bibitem{cordts2016cityscapes}
M.~Cordts, M.~Omran, S.~Ramos, T.~Rehfeld, M.~Enzweiler, R.~Benenson, U.~Franke, S.~Roth, and B.~Schiele, ``The cityscapes dataset for semantic urban scene understanding,'' in \emph{Proceedings of the IEEE conference on computer vision and pattern recognition}, 2016, pp. 3213--3223.

\bibitem{lloyd1982least}
S.~Lloyd, ``Least squares quantization in pcm,'' \emph{IEEE transactions on information theory}, vol.~28, no.~2, pp. 129--137, 1982.

\bibitem{arthur2007k}
D.~Arthur and S.~Vassilvitskii, ``K-means++ the advantages of careful seeding,'' in \emph{Proceedings of the eighteenth annual ACM-SIAM symposium on Discrete algorithms}, 2007, pp. 1027--1035.

\bibitem{shamsian2021personalized}
A.~Shamsian, A.~Navon, E.~Fetaya, and G.~Chechik, ``Personalized federated learning using hypernetworks,'' in \emph{International Conference on Machine Learning}.\hskip 1em plus 0.5em minus 0.4em\relax PMLR, 2021, pp. 9489--9502.

\bibitem{Houben-IJCNN-2013}
S.~Houben, J.~Stallkamp, J.~Salmen, M.~Schlipsing, and C.~Igel, ``Detection of traffic signs in real-world images: The {G}erman {T}raffic {S}ign {D}etection {B}enchmark,'' in \emph{International Joint Conference on Neural Networks}, no. 1288, 2013.

\bibitem{luo2018mio}
Z.~Luo, F.~Branchaud-Charron, C.~Lemaire, J.~Konrad, S.~Li, A.~Mishra, A.~Achkar, J.~Eichel, and P.-M. Jodoin, ``Mio-tcd: A new benchmark dataset for vehicle classification and localization,'' \emph{IEEE Transactions on Image Processing}, vol.~27, no.~10, pp. 5129--5141, 2018.

\bibitem{lecun1989backpropagation}
Y.~LeCun, B.~Boser, J.~S. Denker, D.~Henderson, R.~E. Howard, W.~Hubbard, and L.~D. Jackel, ``Backpropagation applied to handwritten zip code recognition,'' \emph{Neural computation}, vol.~1, no.~4, pp. 541--551, 1989.

\bibitem{he2016deep}
K.~He, X.~Zhang, S.~Ren, and J.~Sun, ``Deep residual learning for image recognition,'' in \emph{Proceedings of the IEEE conference on computer vision and pattern recognition}, 2016, pp. 770--778.

\bibitem{liang2020think}
P.~P. Liang, T.~Liu, L.~Ziyin, N.~B. Allen, R.~P. Auerbach, D.~Brent, R.~Salakhutdinov, and L.-P. Morency, ``Think locally, act globally: Federated learning with local and global representations,'' \emph{arXiv preprint arXiv:2001.01523}, 2020.

\end{thebibliography}
\end{document}